\documentclass[final,3p,Ecuild]{elsarticle}
\usepackage{latexsym}
\usepackage{circledsteps}
\usepackage{amsmath}
\usepackage{pifont}
\usepackage{enumitem} 
\usepackage{amsmath}
\usepackage{amssymb}
\usepackage{amsfonts}
\usepackage{graphicx}  		
\usepackage{color}  		
\usepackage[normalem]{ulem} 
\usepackage{soul} 			
\usepackage{setspace} 		
\usepackage{multirow}		
\usepackage{bm} 			
\usepackage{appendix} 		
\usepackage{lineno} 		
\usepackage{adjustbox}
\usepackage{booktabs}
\usepackage{hyperref}
\usepackage{setspace}

\usepackage{tikz}    
\usetikzlibrary{shapes.geometric, arrows.meta, positioning}

\tikzstyle{startstop} = [rectangle, rounded corners, minimum width=3cm, minimum height=1cm,
                         text centered, draw=black, fill=red!20, font=\large\rmfamily]
\tikzstyle{process} = [rectangle, minimum width=4cm, minimum height=1cm,
                       text centered, draw=black, fill=blue!10, font=\large\rmfamily]
\tikzstyle{decision} = [diamond, aspect=2, minimum width=4cm, minimum height=1cm,
                        text centered, draw=black, fill=green!15, font=\large\rmfamily]
\tikzstyle{arrow} = [semithick , ->, >=Stealth]

\usepackage{multirow}

\usepackage{subfigure}
\usepackage{mathrsfs}

\journal{} 
\begin{document}
\begin{frontmatter}
%
%
\title{Bayesian model updating via streamlined Bayesian active learning cubature}
\address[BTU]{Key Laboratory of Urban Security and Disaster Engineering of Ministry of Education, Beijing University of Technology, Beijing 100124, China}
\address[TUD]{Chair for Reliability Engineering, TU Dortmund University, Leonhard-Euler-Straße 5, Dortmund 44227, Germany}
\address[IJR]{International Joint Research Center for Engineering Reliability and Stochastic Mechanics, Tongji University, Shanghai 200092, China}
\author[BTU]{Pei-Pei Li}
\author[TUD]{Chao Dang\corref{cor1}}
\ead{chao.dang@tu-dortmund.de}
\author[TUD]{Cristóbal H. Acevedo}
\author[TUD]{Marcos A. Valdebenito}
\author[TUD,IJR]{Matthias G.R. Faes}
\cortext[cor1]{Corresponding author.}
%
%
\begin{abstract}\label{sec:abstract}
Bayesian model updating has been widely used for model calibration in engineering and scientific applications, as it not only updates model parameters but also quantifies their uncertainty. 
However, for computationally expensive models, efficiently estimating the model evidence (marginal likelihood) and the posterior distribution remains challenging. 
To address this issue, we propose a novel Bayesian active learning method for Bayesian model updating, which is termed as `Streamlined Bayesian Active Learning Cubature’ (SBALC).
The core idea is to approximate the log-likelihood function using Gaussian process (GP) regression in a streamlined Bayesian active learning way.
Rather than generating many samples from the posterior GP, we only use its mean and variance function to form the model evidence estimator, stopping criterion, and learning function.
Specifically, the estimation of model evidence is first treated as a Bayesian cubature problem, with a GP prior assigned over the log-likelihood function.
Second, a plug-in estimator for model evidence is proposed based on the posterior mean function of the GP.
Third, an upper bound on the expected absolute error between the posterior model evidence and its plug-in estimator is derived.
Building on this result, a novel stopping criterion and learning function are proposed using only the posterior mean and standard deviation functions of the GP.
Finally, we can obtain the model evidence based on the posterior mean function of the log-likelihood function in conjunction with Monte Carlo simulation, as well as the samples for the posterior distribution of model parameters as a by-product.
Four numerical examples are presented to demonstrate the accuracy and efficiency of the proposed method compared to several existing approaches. 
The results show that the method can significantly reduce the number of model evaluations and the computational time without compromising accuracy.

\end{abstract}
\end{frontmatter}
%
%
%
%
%
\section{Introduction}
\label{sec:Introduction}
Computational models are widely used in applied science and engineering to simulate complex systems and support real-world decision making. 
Although they provide valuable insight, their accuracy is highly dependent on the assumptions and data on which they are based.
During system operation and as new measurements are collected, discrepancies often arise between model predictions and observed responses, which result from modeling errors (such as simplifications or neglected physics), parameter uncertainty, measurement errors, and many others.
To address these discrepancies and reduce uncertainty, it is often necessary to update or calibrate uncertain model parameters to enhance predictive accuracy.
Updating these parameters enhances the ability of models to capture real system behavior, improves predictive performance, and supports more reliable decision-making in design, control, and maintenance processes.
Traditional model updating methods, such as deterministic optimization~\cite{friswell1995finite} and gradient-based techniques~\cite{collins1974statistical,mottershead2011sensitivity}, are inherently limited by their deterministic nature. 
They typically yield a single best-fit set of parameters while ignoring uncertainty.  
In recent years, Bayesian model updating has gained great attention for its ability not only to update model parameters but also to quantify the uncertainty associated with them.
In this framework, the prior distribution of model parameters is updated to the posterior distribution based on measurements obtained from the system~\cite{beck1998updating,katafygiotis1998updating}.
Although analytical solutions exist in special cases, such as when conjugate priors are employed, computing the posterior is generally challenging in practical engineering. 
To address this challenge, three main categories of methods are commonly used: approximate analytical methods, simulation-based methods, and surrogate-based methods.

Approximate analytical methods address the Bayesian model updating problem by deriving explicit or semi-explicit expressions for the posterior distribution. 
One representative method is the Laplace approximation~\cite{bleistein1975asymptotic}, which applies a second-order Taylor expansion to the log-posterior distribution around the maximum a posterior (MAP) estimate, yielding a local Gaussian approximation. 
Another important class comprises Kalman filter–based methods.
The classical Kalman filter~\cite{kalman1960new, kalman1961new} obtains the analytical results regarding the Gaussian posterior mean and variance of the model parameters under the assumptions of a linear state transition model, linear observation model, and Gaussian noise.
For nonlinear systems, the Extended Kalman filter~\cite{gelb1974applied,anderson2005optimal} approximates the nonlinear dynamics by applying a first-order Taylor expansion around the current state estimate, while the Unscented Kalman Filter~\cite{julier1997new} avoids explicit linearization by propagating a deterministically chosen set of sigma points through the nonlinear model and estimating the posterior mean and covariance from the transformed points.
In addition to these, variational Bayes~\cite{fox2012tutorial, ni2021probabilistic, jia2023analytically} offers a more general framework by optimizing the evidence lower bound to minimize the Kullback-Leibler divergence between a parameterized variational distribution and the true posterior distribution, thereby yielding a computationally tractable approximation.
While these analytical methods are typically fast and interpretable to some extent, they often rely on restrictive modeling assumptions and local approximations and may perform poorly for strongly nonlinear, non-Gaussian, or multimodal problems.

Instead of pursuing explicit or semi-explicit formulas, simulation-based methods estimate the posterior distribution through numerical simulation, making them highly flexible for complex model updating problems.
A major subclass comprises Markov Chain Monte Carlo (MCMC) algorithms~\cite{metropolis1953equation, hastings1970monte, beck2002bayesian}, which generate dependent samples whose stationary distribution equals the target posterior distribution, using algorithms such as the Metropolis Hastings sampler~\cite{metropolis1953equation,hastings1970monte}, Gibbs sampler~\cite{geman1984stochastic}, and Hamiltonian Monte Carlo~\cite{neal2011mcmc,jensen2025use}. 
However, when the posterior and prior distributions have little overlap, conventional MCMC suffers from low acceptance rates and poor sampling efficiency.
To address this issue, Transitional MCMC (TMCMC)~\cite{ching2007transitional} introduces a sequence of intermediate distributions that gradually transform the prior distribution into the posterior distribution, enabling more efficient exploration of high-probability regions without directly sampling from the sharply peaked posterior.
Another important subclass is particle filters~\cite{gordon1993novel, liu1998sequential}, also known as the Sequential Monte Carlo (SMC).
It approximates the posterior distributions using a set of weighted particles, which are sequentially propagated and updated over time. 
By employing sequential importance sampling and resampling, SMC can effectively handle nonlinear dynamic systems and non-Gaussian noise, making it well-suited for complex model updating problems.
Beyond these two subclasses, the Bayesian Updating with Structural Reliability methods (BUS)~\cite {Straub2015,diazdelao2017bayesian} offer a complementary perspective by reformulating the model updating problem as an equivalent structural reliability analysis.
This framework enables the use of simulation-based reliability methods.
While simulation-based methods are flexible and capable of handling complex posterior distributions, they require a large number of forward model evaluations and are computationally intensive.

Surrogate-based methods have attracted growing interest in recent years due to their ability to balance computational efficiency and estimation accuracy.
These methods replace expensive models with computationally inexpensive approximations, thus improving the efficiency of model updating while maintaining acceptable accuracy.
Among various surrogate modeling techniques such as polynomial chaos expansion (PCE)~\cite{li2024bayesian}, artificial neural networks (ANN)~\cite{atalla1998model}, and Gaussian process regression (GPR) methods~\cite{wan2016stochastic}, the GPR (or Kriging) framework~\cite{seeger2004gaussian} has been extensively studied due to its modeling capability and its ability to provide both predictions and uncertainty quantification.
The first subtype employs an active learning GPR to approximate expensive model evaluations during sampling, thereby improving the efficiency of Bayesian model updating.
Owing to the success of active learning strategies in the reliability analysis field~\cite{echard2011ak,dang2021estimation,dang2022parallel}, the combination of the GPR model with the sampling method within the BUS framework has been widely investigated to improve the computational efficiency~\cite{wang2020highly,liu2022efficient,kitahara2023bayesian,lu2025efficient}.
Beyond the BUS framework, several studies have directly coupled GPR models with simulation-based sampling techniques.
For example, Angelikopoulos et al.~\cite{angelikopoulos2015x} combined adaptive Kriging with TMCMC to reduce the computational cost of approximating the posterior probability distribution.
Yoshida et al.~\cite{yoshida2023bayesian} proposed an adaptive Gaussian process (GP) particle filters framework with a tailored learning function for efficient posterior PDF estimation; and Taflanidis et al.~\cite{taflanidis2025surrogate} embedded an adaptive GPR model within an SMC scheme, using intermediate auxiliary densities to efficiently sample from the target posterior distribution.
The second subtype is GPR-based Bayesian cubature, initially proposed for multidimensional numerical integration~\cite{o1991bayes}.
It was later extended to model evidence estimation~\cite{osborne2012active, gunter2014sampling} by probabilistically integrating the likelihood function, and further to directly approximate the posterior distribution~\cite{wang2018adaptive}.
Model evidence measures the plausibility of the observed data under a specified prior distribution and likelihood function, thus providing a rigorous criterion for model assessment. 
As it is also regarded as an important part in model updating, recent research efforts~\cite{wei2024probabilistic,song2025sampling,kitahara2025sequential} have been focused on its evaluation, with the posterior distribution provided as a by-product.
These studies further demonstrated that coupling Bayesian cubature with active learning for model evidence estimation can substantially reduce the number of model evaluations and yield more robust uncertainty quantification of the approximation results.
However, they typically rely on either generating many sample paths from the GP posterior or on applying a linearization trick to derive closed-form mean and covariance, which is often computationally expensive and cumbersome.
In view of this, it is worthwhile to investigate efficient sample-free Bayesian cubature methods for Bayesian model updating.

In this study, we propose a novel Bayesian active learning method for Bayesian model updating, referred to as `Streamlined Bayesian Active Learning Cubature' (SBALC). 
The proposed method does not require generating samples from the posterior GP, and produces the model evidence and posterior distribution.  
The main contributions of this work can be summarized as follows.
First, we propose a plug-in estimator for the model evidence based on the posterior mean function of the log-likelihood function, which is computationally inexpensive and does not require samples from the posterior GP.
Second, an upper bound on the expected absolute error is derived between the posterior model evidence and its plug-in estimator. 
This enables the quantification of epistemic uncertainty of the plug-in estimator using only the upper and lower credible bounds of the posterior GP.
Third, a novel stopping criterion and a new learning function are defined purely using the posterior mean and variance functions of the GP, providing an efficient and reliable pathway for the active learning process.
Finally, the model evidence can be obtained from the posterior mean function of the GP in conjunction with Monte Carlo simulation (MCS), as well as the posterior distribution of model parameters as a by-product.
It is worth noting that, at the time of submission of this paper, a similar approach has been reported in a preprint by Wei~\cite{wei2025bayesian}.
Nevertheless, the present work differs in both its theoretical derivation and practical implementation.

The remainder of this study is organized as follows.
Section~\ref{sec:Problem formulation} presents the formulation of the problem.
Section~\ref{sec:Streamlined Bayesian active learning cubature} details the proposed SBALC.
Section~\ref{sec:Numerical examples} provides four numerical examples to demonstrate the performance of the proposed method. 
Finally, Section~\ref{sec:Summary_conclusions} summarizes the main conclusions.

%
%
\section{Problem formulation}

\label{sec:Problem formulation}
Bayesian model updating facilitates the calibration of model parameters and the improvement of predictive accuracy of computational models by integrating prior knowledge with newly observed data~\cite{beck1998updating}.
The primary objective of Bayesian model updating is to estimate the posterior probability density function (PDF) of the parameters, conditional on the observed data. 
This objective is achieved through Bayes’ theorem:
\begin{equation} 
f(\boldsymbol{x}\mid\boldsymbol{Y}_{\mathrm{obs}})=\frac{L(\boldsymbol{Y}_{\mathrm{obs}}\mid\boldsymbol{x})f(\boldsymbol{x})}{c},
\label{eq:bayesian_formula}
\end{equation}
where $f(\boldsymbol{x}\mid\boldsymbol{Y}_{\mathrm{obs}})$ denotes the posterior PDF; $\boldsymbol{x}$ denotes a realization of the $d$-dimensional model parameter vector $\boldsymbol{X} = [X_1,X_2,...,X_d]\in D_{\boldsymbol{X}}\subseteq\mathbb{R}^d$ to be updated; $\boldsymbol{Y}_{\mathrm{obs}}$ represents the set of observations obtained from experiments or measurements; $L(\boldsymbol{Y}_{\mathrm{obs}}|\boldsymbol{x})$ is the likelihood function; $f(\boldsymbol{x}): D_{\boldsymbol{X}} \to [0,\infty)$ denotes the prior PDF of the model parameters; and $c$ is the model evidence, also known as the marginal likelihood.

The likelihood function $L(\boldsymbol{Y}_{\mathrm{obs}}|\boldsymbol{x})$ quantifies the probability of observing the data $\boldsymbol{Y}_{\mathrm{obs}}$ given the model parameters $\boldsymbol{x}$ of the computational model $\mathcal{R}(\boldsymbol{X})$.
For simplicity, $\mathcal{R}(\boldsymbol{X})$ is assumed to be a deterministic function, so that all uncertainty arises only from the input model parameters.
The likelihood function typically represents the discrepancy between the response of the computational model $Y=\mathcal{R}(\boldsymbol{X}):\mathbb{R}^d\mapsto\mathbb{R}$ and the observed data $\boldsymbol{Y}_{\mathrm{obs}}$.
If a linear representation of the measurement error is adequate, the error term $\varepsilon$ can be expressed as~\cite{Straub2015,li2023information}:
$\varepsilon=\boldsymbol{Y}_{\mathrm{obs}}-\mathcal{R}(\boldsymbol{X})$.
Alternatively, if the magnitude of the measurement error is expected to scale with the model observations, the error can be expressed in relative terms~\cite{beck2002bayesian}: 
$\varepsilon = \frac{\boldsymbol{Y}{\mathrm{obs}}}{\mathcal{R}(\boldsymbol{X})}-1$.
These two error models, absolute and relative, are commonly used to construct the likelihood function.
The choice of an appropriate error model depends on the nature of the observed data.
To unify these formulations, an error mapping function $\mathcal{J}(\boldsymbol{X})$, defined as $\mathcal{J}(\boldsymbol{X})=J(\boldsymbol{Y}_{\mathrm{obs}},\mathcal{R}(\boldsymbol{X}))$, is introduced.
This function provides a consistent approach for representing errors, regardless of whether absolute or relative errors are considered.
When the PDF of the measurement error is known, the likelihood function can be parameterized by the measurement data and computational model as $L(\boldsymbol{Y}_{\mathrm{obs}}|\boldsymbol{x})=\varphi_{\varepsilon}(\mathcal{J}(\boldsymbol{x}))$,
where $\varphi_{\varepsilon}$ represents the PDF of the measurement error $\varepsilon$.

Various models have been investigated to characterize the PDF of measurement error.
Among these, the Gaussian model is most widely adopted, resulting in the following form of likelihood function:
\begin{equation} \label{eq:likelihood_explicit}
L(\boldsymbol{Y}_{\mathrm{obs}}|\boldsymbol{x})\propto\exp\left(-\frac{\mathcal{J}(\boldsymbol{x})^2}{2\sigma_{\varepsilon}^2}\right),
\end{equation}
where $\sigma_{\varepsilon}$ denotes the standard deviation of the measurement error.

The model evidence  ${c}$  acts as the normalization constant in Bayes’ theorem, ensuring that the posterior distribution integrates into one.
It is defined as the integral of the likelihood function weighted by the prior PDF:
\begin{equation} \label{eq:evidece_formula}
c ={\int_{D_{\boldsymbol{X}}}{L\left(\boldsymbol{Y}_{\mathrm{obs}}|\boldsymbol{x}\right)f\left(\boldsymbol{x}\right)\mathrm{d}\boldsymbol{x}}}.
\end{equation}


Once the prior $f(\boldsymbol{x})$ and likelihood $L(\boldsymbol{Y}_{\mathrm{obs}}|\boldsymbol{x})$ are specified, obtaining the model evidence $c$ and posterior distribution  $f(\boldsymbol{x}\mid\boldsymbol{Y}_{\mathrm{obs}})$ are two key tasks in Bayesian model updating. 
Analytical solutions are typically not available except in special cases where the prior and likelihood are conjugate.
In most engineering applications, however, such conjugacy does not hold, and the underlying models are often nonlinear and computationally expensive to evaluate, necessitating the use of efficient numerical or approximate methods.

\section{Proposed SBALC method}\label{sec:Streamlined Bayesian active learning cubature}
This section introduces the proposed SBALC method for Bayesian model updating, which primarily focuses on model evidence estimation, while also providing samples of posterior distribution as a byproduct.
Section~\ref{subsec:Gaussian process regression} introduces GPR for modeling the log-likelihood function.
Section~\ref{subsec:Plug-in estimator for the model evidence} proposes the plug-in model evidence estimator.
Section~\ref{subsec:Stopping criterion} derives the upper bound of the expected absolute error between the posterior model evidence and its plug-in model evidence estimator, and proposes a novel stopping criterion. 
Section~\ref{subsec:learning_function} presents the new learning function. 
Finally, Section~\ref{subsec:Implementation procedure of the proposed method} outlines the complete implementation procedure of the proposed method.

\subsection{Gaussian process regression}
\label{subsec:Gaussian process regression}
To evaluate the integral in Eq.~\eqref{eq:evidece_formula}, we adopt the Bayesian cubature principle, which interprets the model evidence estimation as a Bayesian inference problem. 
In this setting, the integrand is modeled as a random function through a GP.
However, the likelihood function in Eq.~\eqref{eq:evidece_formula} is strictly non-negative and often exhibits strong nonlinearity, which makes direct modeling difficult.
To address this issue, we instead consider its logarithm, referred to as the log-likelihood function $\mathscr{L}(\boldsymbol{x})=\log(L(\boldsymbol{Y}_{\mathrm{obs}}|\boldsymbol{x}))$$:\mathbb{R}^d\mapsto\mathbb{R}$.
Specifically, a GP prior is placed over $\mathscr{L}(\boldsymbol{x})$:
\begin{equation} 
{\mathscr{L}_{0}}(\omega,\boldsymbol{x})\sim \mathcal{GP}\left(m_{\mathscr{L}_{0}}\left(\boldsymbol{x}\right),\kappa_{\mathscr{L}_{0}}\left(\boldsymbol{x},\boldsymbol{x}^{\prime}\right)\right),
\label{eq:prior_gp}
\end{equation}
where ${\mathscr{L}_0}(\omega,\boldsymbol{x}): \Omega \times D_{\boldsymbol{X}} \to \mathbb{R}$ denotes the prior distribution of $\mathscr{L}(\boldsymbol{x})$ and $\omega\in\Omega$ denotes an elementary event in the probability space $(\Omega, \mathcal{F}, \mathbb{P})$, with $\Omega$ is the sample space, $\mathcal{F}$ is the $\sigma$-algebra of events, and $\mathbb{P}$ is the associated probability measure; $m_{\mathscr{L}_{0}}\left(\boldsymbol{x}\right)$ and $\kappa_{\mathscr{L}_{0}}\left(\boldsymbol{x},\boldsymbol{x}^{\prime}\right)$ denote the prior mean and covariance functions, respectively. 

Without loss of generality, the prior mean can be considered a constant, and the covariance function can be modeled using a Gaussian kernel given by:
\begin{equation} 
m_{\mathscr{L}_{0}}\left(\boldsymbol{x}\right)=\beta,
\label{eq:prior_mean}
\end{equation}
\begin{equation} 
\kappa_{\mathscr{L}_{0}}\left(\boldsymbol{x},\boldsymbol{x}^{\prime}\right)=\sigma_{0}^{2}\;\mathrm{exp}\left(\frac{1}{2}(\boldsymbol{x}-\boldsymbol{x}')^{T}{\boldsymbol \varSigma}^{-1}(\boldsymbol{x}-\boldsymbol{x}')\right),
\label{eq:prior_covariance}
\end{equation}
where $\beta\in\mathbb{R}$; $\sigma_{0}>0$ is the standard deviation of the process; $\boldsymbol \varSigma$ = diag$(l_1^{2},\;l_2^{2},\;...,\;l_d^{2})$ with $l_i>0$ being the length scale in the $i$-th dimension. 
The $d$+2 parameters collected in $\mathcal{V}=[\beta,\sigma_0,l_1,l_2, ...,l_d]$ are referred to as hyperparameters.

Suppose that we have a supervised training dataset $\boldsymbol{\mathcal{D}}=\{\boldsymbol{\mathcal{X}},\boldsymbol{\mathfrak{L}}\}$, where $\boldsymbol{\mathcal{X}}=\{\boldsymbol{x}^{(i)}\}_{i=1}^{n}$ is an $n\times d$ matrix with its $i$-th row being $\boldsymbol{x}^{(i)}$.
$\boldsymbol{\mathfrak{L}} = [\boldsymbol{\mathfrak{L}}^{(1)},\boldsymbol{\mathfrak{L}}^{(2)},...,\boldsymbol{\mathfrak{L}}^{(n)}]^{T}$ contains the corresponding outputs given by the computational model, i.e., $\boldsymbol{\mathfrak{L}}^{(i)} = \mathscr{L}(\boldsymbol{x}^{(i)})$. 
The hyperparameters of the prior mean and covariance functions are estimated via maximum likelihood estimation. 
The posterior of the log-likelihood function conditional on $\boldsymbol{\mathcal{D}}$ remains a GP:
\begin{equation} \mathscr{L}_{n}(\omega,\boldsymbol{x})\sim\mathcal{GP}(m_{\mathscr{L}_{n}}\left(\boldsymbol{x}\right),\kappa_{\mathscr{L}_{n}}\left(\boldsymbol{x},\boldsymbol{x}^{\prime}\right)),
\label{eq:posterior_GP}
\end{equation}
with mean and covariance:
\begin{equation} 
m_{\mathscr{L}_{n}}\left(\boldsymbol{x}\right)=m_{\mathscr{L}_0}\left(\boldsymbol{x}\right)+k_{\mathscr{L}_{0}}(\boldsymbol{x},\boldsymbol{\mathcal{X}})^{T}\boldsymbol{\mathcal{K}}_{\mathscr{L}_{0}}^{-1}\left({\boldsymbol{\mathfrak{L}}}-m_{\mathscr{L}_{0}}(\boldsymbol{\mathcal{X}})\right),
\label{eq:posterior_mean}
\end{equation}
\begin{equation} 
\kappa_{\mathscr{L}_{n}}\left(\boldsymbol{x},\boldsymbol{x}^{\prime}\right)=
\kappa_{\mathscr{L}_{0}}\left(\boldsymbol{x},\boldsymbol{x}^{\prime}\right)
-\kappa_{\mathscr{L}_{0}}\left(\boldsymbol{x},\boldsymbol{\mathcal{X}}\right)^{T}\boldsymbol{\mathcal{K}}_{\mathscr{L}_{0}}^{-1}
\kappa_{\mathscr{L}_{0}}\left(\boldsymbol{\mathcal{X}},\boldsymbol{x}^{\prime}\right),
\label{eq:posterior_covariance}
\end{equation}
where $m_{\mathscr{L}_{0}}(\boldsymbol{\mathcal{X}})$ is an $n\times1$ mean vector with $i$-th element being $m_{\mathscr{L}_{0}}(\boldsymbol{x}^{(i)})$; $\boldsymbol{\mathcal{K}}_{\mathscr{L}_{0}}$ is an $n$-by-$n$ a covariance matrix whose $(i_1,i_2)$-th element is $\kappa_{\mathscr{L}_{0}}\left(\boldsymbol{x}^{(i_1)},\boldsymbol{x}^{(i_2)}\right)$ at the $i_1$-th and $i_2$-th training samples;
$\kappa_{\mathscr{L}_{0}}\left(\boldsymbol{x}, \boldsymbol{\mathcal{X}}\right)$ and $\kappa_{\mathscr{L}_{0}}\left(\boldsymbol{\mathcal{X}},\boldsymbol{x}^{\prime}\right)$ are two $n\times1$ covariance vectors with $i$-th element being $\kappa_{\mathscr{L}_{0}}\left(\boldsymbol{x},\boldsymbol{x}^{(i)}\right)$ and $\kappa_{\mathscr{L}_{0}}\left(\boldsymbol{x}^{(i)},\boldsymbol{x}^{\prime}\right)$, respectively. 

The posterior GP provides not only a mean prediction, but also a measure of uncertainty in the prediction at any unseen point, given by the posterior variance $\sigma^2_{\mathscr{L}_{n}}(\boldsymbol{x})=\kappa_{\mathscr{L}_n}\left(\boldsymbol{x},\boldsymbol{x}\right)$.

\subsection{Plug-in estimator for the model evidence}
\label{subsec:Plug-in estimator for the model evidence}
Since the $\mathscr{L}_{n}(\omega,\boldsymbol{x})$ follows a Gaussian process, exponentiating it yields the posterior distribution of the likelihood function $e^{\mathscr{L}_{n}(\omega,\boldsymbol{x})}$, which follows a lognormal process.
The posterior model evidence is defined as: 
\begin{equation} 
{c_n}(\omega) ={\int_{D_{\boldsymbol{X}}}{e^{\mathscr{L}_{n}(\omega,\boldsymbol{x})}f\left(\boldsymbol{x}\right)\mathrm{d}\boldsymbol{x}}}.
\label{eq:evidence_estimator}
\end{equation}

Since the posterior model evidence is a function of $\omega$, it can be regarded as a random variable.
The randomness of posterior model evidence originates from the GP model itself, reflecting epistemic uncertainty due to the limited number of log-likelihood evaluations. 
As more evaluations of the log-likelihood function are collected, this uncertainty may diminish and the posterior model evidence $c_n(\omega)$ converges to the true model evidence $c^\ast$. 
However, since the exact distribution of $c_n(\omega)$ is analytically intractable, we introduce a plug-in model evidence estimator by replacing $\mathscr{L}_{n}(\omega,\boldsymbol{x})$ with its posterior mean function:
\begin{equation} 
{\tilde{c}_n} ={\int_{D_{\boldsymbol{X}}}{e^{m_{\mathscr{L}_{n}}(\boldsymbol{x})}f\left(\boldsymbol{x}\right)\mathrm{d}\boldsymbol{x}}}.
\label{eq:evidence_mean_esimator}
\end{equation}

This plug-in estimator will also converge to the true model evidence $c^\ast$ as more evaluations of the log-likelihood function become available (when $ \sigma_{\mathscr{L}_{n}}(\boldsymbol{x}) \rightarrow 0$ and $m_{\mathscr{L}_{n}}(\boldsymbol{x}) \rightarrow \mathscr{L}(\boldsymbol{x})$ for  $ \forall \boldsymbol{x} \in D_{\boldsymbol{x}}$).
However, the integral in Eq.~\eqref{eq:evidence_mean_esimator} is generally intractable in closed form.
Therefore, Monte Carlo simulation (MCS) is adopted to provide the numerical estimate of the plug-in model evidence $\tilde{c}_n$.
The MCS estimator of $\tilde{c}_n$ is given by:
\begin{equation} 
\hat{\tilde{c}}_n =\frac{1}{N_{M}}\sum_{j=1}^{N_{M}} {e^{m_{\mathscr{L}_{n}}(\boldsymbol{x}^{(j)})}},
\label{eq:evidence_mean_mcs}
\end{equation}
where $\boldsymbol{x}^{(j)}$ is the $j$-th sample in the sample set $\mathbb{S}={\{\boldsymbol{x}^{(1)},...,\boldsymbol{x}^{(N_M)}}\}$ drawn from the prior PDF; $N_{M}$ denotes the number of random samples.

The variance of the estimator $\hat{\tilde{c}}_n$ is expressed as:
\begin{equation}
\mathrm{VAR}\big[\hat{\tilde{c}}_n
\big] 
=\frac{1}{N_M(N_M-1)} \sum\limits_{j=1}^{N_M} 
\left(e^{m_{\mathscr{L}_{n}}(\boldsymbol{x}^{(j)})} - {\hat{\tilde{c}}_n}\right)^2,
\label{eq:sd_estimator}
\end{equation}
where $\mathrm{VAR}$ denotes the variance operator.

Furthermore, the coefficient of variation (CoV) of the estimator $\hat{\tilde{c}}_n$ is defined as $\mathrm{CoV}\big[\hat{\tilde{c}}_n\big]=\sqrt{\mathrm{VAR}[\hat{\tilde{c}}_n]}\big/\hat{\tilde{c}}_n$, which scales as $\mathcal{O}(1/\sqrt{N_M})$.
This indicates that a large sample size leads to a more reliable estimate.

\subsection{Uncertainty measure and stopping criterion}
\label{subsec:Stopping criterion}
After obtaining the model evidence estimate, a stopping criterion is needed to evaluate whether the estimate reaches a desired level of accuracy.
Based on the predictions of the posterior GP for the log-likelihood function, the upper credible bound and the lower credible bound of the posterior GP are defined as $u_{n}(\boldsymbol{x}) = m_{\mathscr{L}_{n}}(\boldsymbol{x})+b\sigma_{\mathscr{L}_{n}}(\boldsymbol{x})$ and
 $l_{n}(\boldsymbol{x}) = m_{\mathscr{L}_{n}}(\boldsymbol{x})-b\sigma_{\mathscr{L}_{n}}(\boldsymbol{x})$. 
Here, the parameter $b$ ($b>0$) is the real number controls the width of the credible interval; in particular, setting $b = \Phi^{-1}(1-\alpha/2)$ yields a $(1-\alpha)\times100\%$ credible interval, where $\Phi^{-1}(\cdot)$ denotes the inverse cumulative distribution function (CDF) of the standard normal variable.
If the posterior distribution $\mathscr{L}_{n}(\omega,\boldsymbol{x})$ in Eq.~\eqref{eq:evidence_estimator} is replaced by its upper and lower credible bounds, two new quantities can be defined:
\begin{equation}
\overline{c}_n   = \int_{D_{\boldsymbol{X}}}\left(e^{(m_{\mathscr{L}_{n}}(\boldsymbol{x})+b\sigma_{\mathscr{L_{n}}}(\boldsymbol{x}))}\right) f\left(\boldsymbol{x}\right)\mathrm{d}\boldsymbol{x},
\label{eq:evidence_lower}
\end{equation}
\begin{equation} 
\underline{c}_n =\int_{D_{\boldsymbol{X}}}\left(e^{(m_{\mathscr{L}_{n}}(\boldsymbol{x})-b\sigma_{\mathscr{L_{n}}}(\boldsymbol{x}))}\right) f\left(\boldsymbol{x}\right)\mathrm{d}\boldsymbol{x},
\label{eq:evidence_upper}
\end{equation}
where $\overline{c}_n$ and $\underline{c}_n$ are the upper and lower bounds of the posterior model evidence $c_n(\omega)$.
These bounds cannot be regarded as $(1-\alpha)\times100\%$ credible interval for $c_n(\omega)$, because the integration operator does not, in general, preserve the coverage probability implied by the point-wise credible bounds of $\mathscr{L}_{n}(\omega,\boldsymbol{x})$.

\textbf{Proposition 1.} 
For $b>0$, there exists $\underline{c}_n \leq \tilde{c}_n \leq \overline{c}_n$.

\textbf{Proof.}

Based on Eqs.\eqref{eq:evidence_mean_esimator} and \eqref{eq:evidence_lower}, we have:
\begin{equation} 
\underline{c}_n-\tilde{c}_n =
\int_{D_{\boldsymbol{X}}}
\left[ e^{-b \sigma_{\mathscr{L}_n}(\boldsymbol{x})} - 1 \right]
e^{m_{\mathscr{L}_n}(\boldsymbol{x})} f(\boldsymbol{x})
\,\mathrm{d}\boldsymbol{x}.
\label{eq:evidence_inequlity_factorize}
\end{equation}

Since $b>0$ and $\sigma_{\mathscr{L}_n}(\boldsymbol{x}) \ge 0$, we have $e^{-b\,\sigma_{\mathscr{L}_n}(\boldsymbol{x})} - 1 \le 0$, with equality if and only if $\sigma_{\mathscr{L}_n}(\boldsymbol{x}) = 0$.
Multiplying this inequality by the positive function $e^{\,m_{\mathscr{L}_n}(\boldsymbol{x})} f(\boldsymbol{x})$ preserves the sign, and integration over the input space yields $\underline{c}_n \le \tilde{c}_n$, with equality only when $\sigma_{\mathscr{L}_n}(\boldsymbol{x}) = 0$.
Similarly to the proof of the first inequality $\underline{c}_n \le \tilde{c}_n$, the second inequality 
$\tilde{c}_n \le \overline{c}_n$ can also be proved. 
Combining $\underline{c}_n \le \tilde{c}_n$ and $ \tilde{c}_n \le \overline{c}_n$ completes the proof, holding if and only if $\sigma_{\mathscr{L}_n}(\boldsymbol{x}) = 0$.

\textbf{Proposition 1} indicates that for any $b>0$, the plug-in model evidence estimator $\tilde{c}_n$ always lies between its lower and upper bounds. 
Moreover, as the standard deviation $\sigma_{\mathscr{L}_n}(\boldsymbol{x})$ of the posterior GP tends to zero, both bounds converge to the plug-in model evidence estimator: $\underline{c}_n\rightarrow{\tilde{c}_n}$ and $\overline{c}_n\rightarrow{\tilde{c}_n}$.

\textbf{Proposition 2.} 
For $b>0$, there exists a finite constant $M_b$ such that $\mathbb{E}[|c_n(\omega)-\tilde{c}_n|]\leq  M_b\cdot(\overline{c}_n-\underline{c}_n)$.

\textbf{Proof.}

From Eqs.~\eqref{eq:evidence_estimator} and ~\eqref{eq:evidence_mean_esimator}, the expected absolute error between $c_n(\omega)$ and $\tilde{c}_n$ is given by: 
\begin{equation} 
\mathbb{E}_{\omega}[|c_n(\omega)-\tilde{c}_n|] = \mathbb{E}_{\omega}\left[\left |\int_{D_{\boldsymbol{X}}}{e^{{\mathscr{L}_n}(\omega,\boldsymbol{x})}-e^{{m_{\mathscr{L}_n}}(\boldsymbol{x})}f(\boldsymbol{x})d\boldsymbol{x}}\right|\right].
\label{eq:error_defination}
\end{equation}
where $E_{{\omega}}[\cdot]$ denotes the expectation over $\omega$.

By the triangle inequality and Tonelli's theorem, Eq.~\eqref{eq:error_defination} can be expressed as:

\begin{equation} 
\mathbb{E}_{\omega}[|c_n(\omega)-\tilde{c}_n| ]\leq \int_{D_{\boldsymbol{X}}} \mathbb{E}_{\omega}\left[\left|e^{\mathscr{L}_{n}(\omega,\,\boldsymbol{x})}-{e^{m_{\mathscr{L}_{n}}(\boldsymbol{x})}}\right|\right]f(\boldsymbol{x})d\boldsymbol{x}.
\label{eq:error_inequality}
\end{equation}

To evaluate the expectation in the integrand of Eq.~\eqref{eq:error_inequality}, we fix $\boldsymbol x^\ast\in D_{\boldsymbol X}$.
By the properties of GPR, $\mathscr{L}_{n}(\omega,\,\boldsymbol{x^*})$ follows a Gaussian distribution:
\begin{equation} 
\mathscr{L}_{n}(\omega,\boldsymbol x^*) \overset{d}{=} m_{\mathscr{L}_{n}}(\boldsymbol x^*) + \sigma_{\mathscr{L}_{n}}(\boldsymbol x^*) Z,\quad Z \sim \mathcal{N}(0,1).
\label{eq:GP_ marginal}
\end{equation}

Hence, the expectation in the integrand of Eq.~\eqref{eq:error_inequality} can be reformulated in terms of $Z$:
\begin{equation}
\mathbb E_\omega\!\left[\left|e^{\mathscr L_{n}(\omega,\boldsymbol x^*)}-e^{m_{\mathscr L_{n}}(\boldsymbol x^*)}\right|\right]
= e^{\,m_{\mathscr L_{n}}(\boldsymbol x^*)}\,
\mathbb E_Z\!\left[\left|e^{\sigma_{\mathscr L_{n}}(\boldsymbol x^*)\,Z}-1\right|\right].
\label{eq:mt-factorize}
\end{equation}

Let $a:=\sigma_{\mathscr L_{n}}(\boldsymbol x^*)\ge0$.
The remaining expectation in Eq.~\eqref{eq:mt-factorize} can be evaluated as:
\begin{equation}
\mathbb E_Z\!\left[|e^{aZ}-1|\right]
=\int_{0}^{\infty}(e^{az}-1)\phi(z)\,dz+\int_{-\infty}^{0}(1-e^{az})\phi(z)\,dz,
\label{eq:split}
\end{equation}
where $\phi(z)=\frac{1}{\sqrt{2\pi}}e^{-z^2/2}$ denotes the  PDF of $Z$. 

Applying the change of variable $y=-z$ to the second integral in Eq.~\eqref{eq:split}, it follows that
\begin{equation}
\int_{-\infty}^{0}(1-e^{az})\phi(z)\,dz=\int_{0}^{\infty}\bigl(1-e^{-ay}\bigr)\phi(y)\,dy.
\label{eq:change}
\end{equation}

By substituting Eq.~\eqref{eq:change} into Eq.~\eqref{eq:split} and using the relation $\int_{0}^{\infty} \phi(z)\, dz \;=\frac{1}{2}$, we obtain:
\begin{equation}
\mathbb E_Z\!\left[|e^{aZ}-1|\right]
=\underbrace{\int_0^\infty e^{az}\phi(z)\,dz}_{=:I_+(a)}\;-\;\underbrace{\int_0^\infty e^{-az}\phi(z)\,dz}_{=:I_-(a)},
\label{eq:Iplusminus}
\end{equation}
where the integrals $I_+(a)$ and $I_-(a)$ can be evaluated explicitly as follows:
\begin{equation}
I_+(a)=\frac{1}{\sqrt{2\pi}}\int_0^\infty e^{-(z^2/2)+az}\,dz
=\frac{1}{\sqrt{2\pi}}\int_0^\infty e^{-\frac{(z-a)^2}{2}+\frac{a^2}{2}}\,dz
=e^{a^2/2}\,\Phi(a),
\label{eq:Iplusminus_plus}
\end{equation}
\begin{equation}
I_-(a)=\frac{1}{\sqrt{2\pi}}\int_0^\infty e^{-(z^2/2)-az}\,dz
=e^{a^2/2}\,\Phi(-a).
\label{eq:Iplusminus_minus}
\end{equation}

By substituting Eqs.~\eqref{eq:Iplusminus_plus} and ~\eqref{eq:Iplusminus_minus} into Eq.~\eqref{eq:Iplusminus}, we obtain:
\begin{equation}
\mathbb E_Z\!\left[|e^{aZ}-1|\right]=e^{a^2/2}\bigl(\Phi(a)-\Phi(-a)\bigr)
=e^{a^2/2}\bigl(2\Phi(a)-1\bigr).
\label{eq:Iplusminus_proof}
\end{equation}

Inserting Eq.~\eqref{eq:Iplusminus_proof} into Eq.~\eqref{eq:mt-factorize}, and further into Eq.~\eqref{eq:error_inequality}, leads to:
\begin{equation}
\mathbb{E}_\omega\!\left[\left|c_n(\omega)-\tilde{c}_n \right| \right]
\le \int_{D_{\boldsymbol{X}}} 
e^{\,m_{\mathscr{L}_{n}}(\boldsymbol{x})+\frac{1}{2}\sigma_{\mathscr L_{n}}^2(\boldsymbol{x})}
\big[\,2\Phi(\sigma_{\mathscr L_{n}}(\boldsymbol{x}))-1\,\big]\,
f(\boldsymbol{x})\,d\boldsymbol{x},
\label{eq:error_derive_total}
\end{equation}

On the other hand, the difference between $u_n(\boldsymbol{x})$ and $l_n(\boldsymbol{x})$ is given as:
\begin{equation}
u_n(\boldsymbol{x})-l_n(\boldsymbol{x})=e^{m_{\mathscr{L}_n}(\boldsymbol{x})}\left( e^{b\sigma_{\mathscr{L}_n}(\boldsymbol{x})}-e^{-b \sigma_{\mathscr{L}_n}(\boldsymbol{x})} \right)=2e^{m_{\mathscr{L}_{n}}(\boldsymbol{x})}\text{sinh}\left(b\sigma_{\mathscr{L}_{n}}(\boldsymbol{x})\right).
\label{eq:difference_upper_lower}
\end{equation}

By rearranging Eq.~\eqref{eq:difference_upper_lower}, we can isolate $e^{m_{\mathscr{L}_{n}}(\boldsymbol{x})}$ as
\begin{equation}
e^{m_{\mathscr{L}_{n}}(\boldsymbol{x})} = \frac{u_n(\boldsymbol{x})-l_n(\boldsymbol{x})}{2 \sinh(b \sigma_{\mathscr{L}_{n}}(\boldsymbol{x}))}.
\label{eq:definition_mean_predictor}
\end{equation}

Inserting Eq.~\eqref{eq:definition_mean_predictor} into Eq.~\eqref{eq:error_derive_total} yields:
\begin{equation}
\mathbb{E}_\omega\!\left[\left|c_n(\omega)-\tilde{c}_n \right| \right]
\le \int_{D_{\boldsymbol{X}}}\left(u_n(\boldsymbol{x})-l_n(\boldsymbol{x})\right)h_b(\sigma_{\mathscr L_{n}}(\boldsymbol{x}))
f(\boldsymbol{x})d\boldsymbol{x},
\label{eq:error_bounds_define}
\end{equation}
where
\begin{equation}
h_b(\sigma_{\mathscr L_{n}}(\boldsymbol{x}))=\frac{
e^{\frac{1}{2}\sigma_{\mathscr L_{n}}^2(\boldsymbol{x})}
\big[\,2\Phi(\sigma_{\mathscr L_{n}}(\boldsymbol{x}))-1\,\big]\,}{2 \sinh(b \sigma_{\mathscr{L}_{n}}(\boldsymbol{x}))}.
\label{eq:hb_formula}
\end{equation}

For the Gaussian kernel, there exists $\sigma_0>0$ such that $\sigma_{\mathscr L_{n}}(\boldsymbol{x})\in[0,\sigma_0]$ for all $\boldsymbol{x}$.
Consider the function $h_b(\sigma_{\mathscr L_{n}}(\boldsymbol{x}))$ on the interval $[0,\sigma_0]$.
By applying the inequality $2\Phi(t)-1 \le \min\{\sqrt{2/\pi}\,t,\, 1\}$ for $t\,\ge 0$ and the bound $\sinh(z) \ge z$ for $(z\,\ge0)$, we obtain:

\begin{equation}
h_b(\sigma_{\mathscr L_{n}}(\boldsymbol{x})) \le 
\begin{cases}
\displaystyle \tfrac{e^{1/2}}{2b}\sqrt{\tfrac{2}{\pi}}, & 0 \le \sigma_{\mathscr L_{n}}(\boldsymbol{x}) \le 1, \\[1.2ex]
\displaystyle \tfrac{e^{\sigma_0^2/2}}{2\sinh(b)}, & 1 \le \sigma_{\mathscr L_{n}}(\boldsymbol{x}) \le \sigma_0.
\end{cases}
\label{eq:hb_formula_cases}
\end{equation}

By taking the supremum of $h_b(\sigma_{\mathscr L_{n}}(\boldsymbol{x}))$ with respect to $\sigma_{\mathscr L_{n}}(\boldsymbol{x})\in[0,\sigma_0]$ in Eq.~\eqref{eq:hb_formula_cases}, we obtain
\begin{equation}
M_b := \sup_{\sigma_{\mathscr L_{n}}(\boldsymbol{x}) \in [0,\sigma_0]} h_b(\sigma_{\mathscr L_{n}}(\boldsymbol{x})) 
\ \le \ \max\left\{ \frac{e^{1/2}}{2b} \sqrt{\frac{2}{\pi}}, \ \frac{e^{\sigma_0^2/2}}{2\sinh(b)} \right\}.
\label{eq:Mb_formula}
\end{equation} 

Moreover, using the first-order expansions $\Phi(t)=\tfrac12+\tfrac{t}{\sqrt{2\pi}}+o(t)$ and $\sinh(z)=z+o(z)$ as $t,z\to 0$, we have
$\lim_{\sigma_{\mathscr L_{n}}(\boldsymbol{x})\to 0}
h_b\bigl(\sigma_{\mathscr L_{n}}(\boldsymbol{x})\bigr)
=\lim_{s\to 0}\frac{e^{s^{2}/2}\bigl(2\Phi(s)-1\bigr)}{2\sinh(bs)}
=\frac{\sqrt{2/\pi}}{2b}$.
Hence $h_b$ admits a continuous extension at $0$ and is bounded on $[0,\sigma_0]$; let 
$M_b:=\sup_{\sigma_{\mathscr L_{n}}(\boldsymbol{x})\in[0,\sigma_0]} h_b(\sigma_{\mathscr L_{n}}(\boldsymbol{x}))<\infty$.
Therefore, 
\begin{equation}
\mathbb{E}_\omega\!\left[\,|c_n(\omega)-\tilde{c}_n|\,\right] \le M_b \int_{D_{\boldsymbol{X}}} \left(e^{m_{\mathscr{L}_n}(\boldsymbol{x})+b\sigma_{\mathscr{L}_n}(\boldsymbol{x})}-e^{m_{\mathscr{L}_n}(\boldsymbol{x})-b \sigma_{\mathscr{L}_n}(\boldsymbol{x})} \right)f(\boldsymbol{x}) \, d\boldsymbol{x} 
= M_b\cdot(\overline{c}_n-\underline{c}_n).
\label{eq:error_final}
\end{equation} 

\textbf{Proposition 2} shows that the expected absolute error between $c_n(\omega)$ and $\tilde{c}_n$ is bounded by a constant multiple of the difference between the upper and lower bounds of $c_n(\omega)$, namely $\overline{c}_n-\underline{c}_n$.
This result implies that the epistemic uncertainty in ${c}_n(\omega)$ can be quantified solely through $\overline{c}_n-\underline{c}_n$.
Moreover, if $\overline{c}_n-\underline{c}_n$ converges to zero, then the expected absolute error $\mathbb{E}_\omega[|c_n(\omega)-\tilde{c}_n|]$ also converges to zero, thereby providing a theoretical guarantee that the plug-in model evidence estimator converges to the true model evidence $c^*$.

Taken together, Propositions \textbf{1} and \textbf{2} provide a rigorous theoretical foundation for designing the stopping criterion and learning function.

Based on the above prepositions, a stopping criterion is proposed as follows:
\begin{equation}
\frac{\overline{c}_n -\underline{c}_n}{\tilde{c}_n} < \epsilon_{\text{RE}},
\label{eq:stopping_cri}
\end{equation}
where $\epsilon_\text{RE}$ denotes the user-defined threshold of the relative error. 

The stopping criterion in Eq.~\eqref{eq:stopping_cri} requires that the difference between the upper and lower bounds of the posterior model evidence $\overline{c}_n-\underline{c}_n$ relative to the plug-in estimator $\tilde{c}_n$ needs to be smaller than a specified threshold $\epsilon_{\text{RE}}$.
In other words, the iterative procedure ends once the relative epistemic uncertainty in ${c}_n(\omega)$ is sufficiently small.
To implement the proposed stopping criterion, all components in Eq.~\eqref{eq:stopping_cri} need to be estimated. 
The denominator $\tilde{c}_n$ can be computed using MCS as shown in Eq.~\eqref{eq:evidence_mean_mcs}. 
Likewise, the $\overline{c}_n$ and $\underline{c}_n$ in the numerator can also be estimated via MCS, and are given by:
\begin{equation} 
\hat{\overline{c}}_n =\frac{1}{N_{M}}\sum_{j=1}^{N_{M}} {e^{(m_{\mathscr{L}_n}(\boldsymbol{x}^{(j)})+b\sigma_{\mathscr{L_n}}(\boldsymbol{x}^{(j)}))}},
\label{eq:evidence_upper_mcs}
\end{equation}
\begin{equation} 
\hat{\underline{c}}_n =\frac{1}{N_{M}}\sum_{j=1}^{N_{M}} {e^{(m_{\mathscr{L}_n}(\boldsymbol{x}^{(j)})-b\sigma_{\mathscr{L}_n}(\boldsymbol{x}^{(j)}))}}.
\label{eq:evidence_lower_mcs}
\end{equation}

\textbf{Remark 1.} In addition to the stopping criterion in Eq.~\eqref{eq:stopping_cri}, other two possible options are:  $\frac{\overline{c}_n -\tilde{c}_n}{\tilde{c}_n} < \epsilon_{\text{RE}}$ and  $\frac{\tilde{c}_n-\underline{c}_n}{\tilde{c}_n} < \epsilon_{\text{RE}}$.

\subsection{Learning function and the best next point selection}
\label{subsec:learning_function}
If the stopping criterion has not yet been met, a well-designed learning function is needed to select the best next point to improve the quality of the GPR surrogate model of the log-likelihood function, aiming to further improve the accuracy of the model evidence estimate.
For this purpose, we introduce a new learning function defined as:
\begin{equation} 
\mathcal{LF}(\boldsymbol{x})=\underbrace{\sigma^2_{\mathscr{L}_n}(\boldsymbol{x})}_{\Circled{1}}
\cdot\underbrace{\left(e^{m_{\mathscr{L}_n}(\boldsymbol{x})+b\sigma_{\mathscr{L}_n}(\boldsymbol{x})}-e^{m_{\mathscr{L}_n}(\boldsymbol{x})-b\sigma_{\mathscr{L}_n}(\boldsymbol{x})}\right)\cdot{f(\boldsymbol{x})}}_{\Circled{2}},
\end{equation}\label{eq:learning_function}
where the first term ${\Circled{1}}$ serves as a weighting factor that emphasizes regions of high uncertainty, while the second term ${\Circled{2}}$ contributes to the difference between the upper and lower bounds of the posterior model evidence.
Actually, it is the integrand of $\overline{c}_n -\underline{c}_n$.
A larger value of $\mathcal{LF}(\boldsymbol{x})$ indicates that the candidate point is more promising.

The next evaluation point is selected by maximizing the learning function:
\begin{equation} 
\boldsymbol{x}^{(n+1)} 
= \arg\max_{\boldsymbol{x}\in D_{\delta_1}} \mathcal{LF}(\boldsymbol{x}),
\label{eq:best_point}
\end{equation}
where the feasible domain $D_{\delta_1}$ is defined as 
$D_{\delta_1}=\prod_{j=1}^d \big[ F_j^{-1}(\delta_1), \; F_j^{-1}(1-\delta_1) \big]$, with $F_j^{-1}$ is the inverse prior CDF of $X_j$ and $\delta_1$ is a small truncation probability.

\textbf{Remark 2.} If $\frac{\overline{c}_n -\tilde{c}_n}{\tilde{c}_n} < \epsilon_{\text{RE}}$ is adopted as the stopping criterion, the corresponding learning function can be:  $\mathcal{LF}(\boldsymbol{x})= \sigma^2_{\mathscr{L}_n}(\boldsymbol{x})
\cdot\left(e^{m_{\mathscr{L}_n}(\boldsymbol{x})+b\sigma_{\mathscr{L}_n}(\boldsymbol{x})}-e^{m_{\mathscr{L}_n}(\boldsymbol{x})}\right)\cdot{f(\boldsymbol{x})}$.

\textbf{Remark 3.} If $\frac{\tilde{c}_n - \underline{c}_n}{\tilde{c}_n} < \epsilon_{\text{RE}}$ is adopted as the stopping criterion, the corresponding learning function can be:  $\mathcal{LF}(\boldsymbol{x})= \sigma^2_{\mathscr{L}_n}(\boldsymbol{x})
\cdot\left(e^{m_{\mathscr{L}_n}(\boldsymbol{x})}-e^{m_{\mathscr{L}_n}(\boldsymbol{x}) - b\sigma_{\mathscr{L}_n}(\boldsymbol{x}) }\right)\cdot{f(\boldsymbol{x})}$.

\subsection{Implementation procedure of the proposed method}
\label{subsec:Implementation procedure of the proposed method}
\begin{figure}[htbp]
\centering
\begin{tikzpicture}[node distance=1.8cm and 0.6cm, scale=0.6, transform shape]

\node (start) [startstop] {Start};

\node (step1) [process, below of=start] {Define $f(\boldsymbol{x})$, $\boldsymbol{Y}_{\mathrm{obs}}$, and  $L(\boldsymbol{x})$.};

\node (step2) [process, below of=step1] {Define an initial sampling pool: $\mathbb{S} = \{\boldsymbol{x}^{(j)}\}_{j=1}^{N_M}$};

\node (step3) [process, below of=step2] {Generate an initial training dataset: $\boldsymbol{\mathcal{D}}= \{\boldsymbol{\mathcal{X}}, \boldsymbol{\mathfrak{L}}\}$ and let $n=n_0$};

\node (step4) [process, below of=step3] {Build a GPR model $\mathscr{L}_n(\boldsymbol{x})$ based on $\boldsymbol{\mathcal{D}}$};

\node (step5) [process, below of=step4] {Estimate $\hat{\tilde{c}}_n$, $\hat{\overline{c}}_n$, $\hat{\underline{c}}_n$};

\node (dec1) [decision, below of=step5, yshift=-1cm] {Stopping criterion \#1?};

\node (dec2) [decision, below of=dec1, yshift=-1.5cm] {Stopping criterion \#2?};

\node (step8) [process, below of=dec2, yshift=-1cm] {Approximate the posterior distribution $\hat{f}(\boldsymbol{x}|\boldsymbol{Y}_{\mathrm{obs}})$};

\node (stop) [startstop, below of=step8] {Stop};

\node (enrichTrain) [process, right=2cm of dec1] {Enrich the training dataset $\boldsymbol{\mathcal{D}}= \boldsymbol{\mathcal{D}}\cup\{\boldsymbol{x}^{(n+1)},\boldsymbol{\mathfrak{L}}^{(n+1)}\}$ };

\node (enrichPool) [process, left=2cm of dec2] {Enrich the sample pool $\mathbb{S}=\mathbb{S}\cup{\mathbb{S}^+}$};

\draw [arrow] (start) -- (step1);
\draw [arrow] (step1) -- (step2);
\draw [arrow] (step2) -- (step3);
\draw [arrow] (step3) -- (step4);
\draw [arrow] (step4) -- (step5);
\draw [arrow] (step5) -- (dec1);

\draw [arrow] (dec1.east) -- node[above]{No} (enrichTrain.west);
\draw [arrow] (dec1) -- node[right]{Yes} (dec2);
\draw [arrow] (dec2.west) -- node[above]{No} (enrichPool.east);
\draw [arrow] (enrichPool.north) |- node[pos=0.25, left] {$ N_M = N_M + N$}(step5.west); 
\draw [arrow] (enrichTrain.north) |- node[pos=0.25, right] {\large$ n = n + 1$} (step4.east);

\draw [arrow] (dec2) -- node[right]{Yes} (step8);
\draw [arrow] (step8) -- (stop);

\end{tikzpicture}
 \caption{Flowchart of the proposed method}
    \label{fig:proposed_flowchart}
\end{figure}
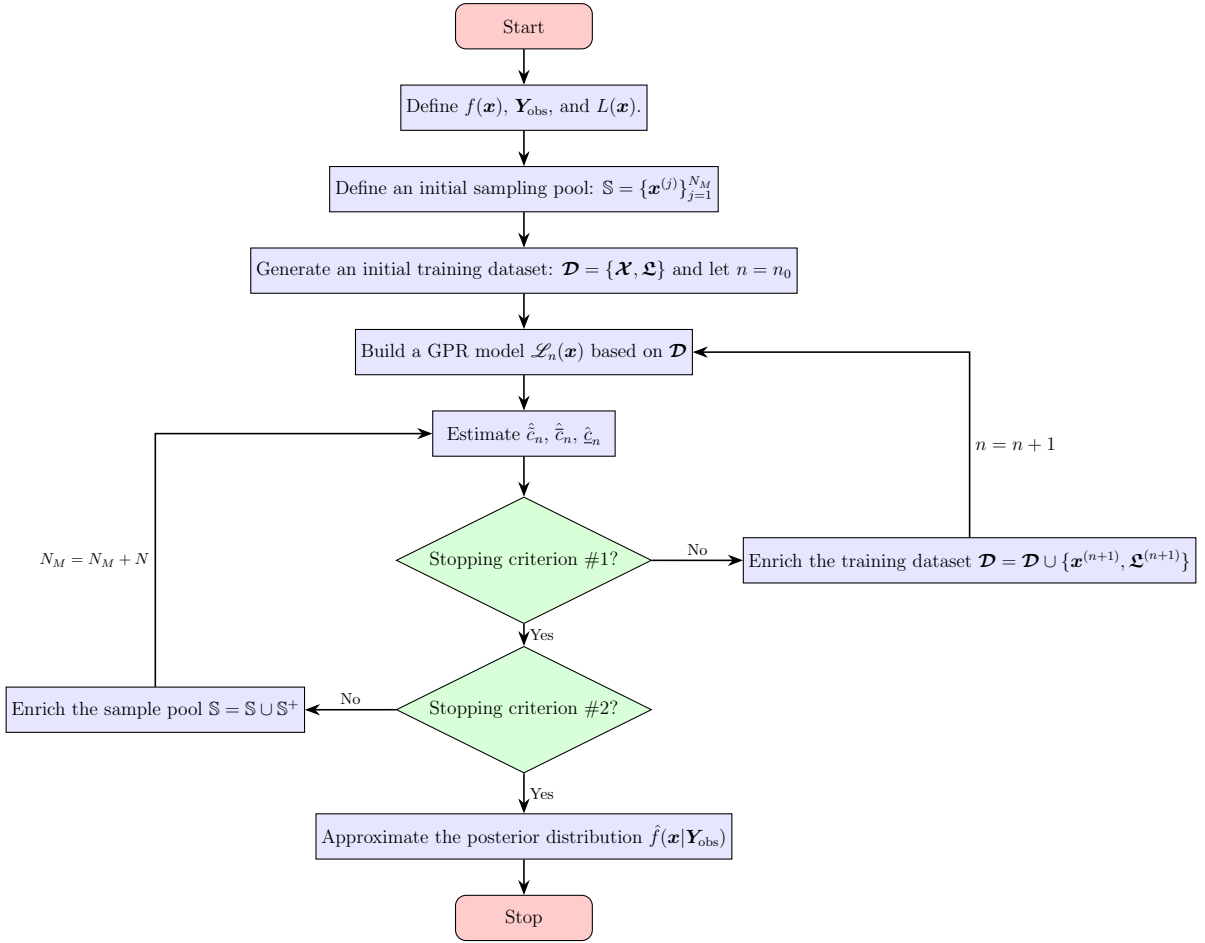

The implementation procedure of the proposed SBALC method is summarized below, along with a flowchart in Fig.~\ref{fig:proposed_flowchart}.

\textbf{Step 1: Initialization}

Define the prior distribution $f(\boldsymbol{x})$ of model parameters, and specify the likelihood function $L(\boldsymbol{x})$ based on observed data $\boldsymbol{Y}_{\mathrm{obs}}$, the specified measurement error ${\varepsilon}$, and the response function $\mathcal{R}(\boldsymbol{X})$. 

\textbf{Step 2: Generate prior samples} 

Generate a set of $N_M$ random samples according to  the prior PDF $f(\boldsymbol{x})$, denoted as $\mathbb{S}={\{\boldsymbol{x}^{(j)}}\}$ for $j = 1,2,...,N_M$.

\textbf{Step 3: Generate an initial training dataset}

First, a small number of $n_0$ input samples $\boldsymbol{\mathcal{X}} = \left\{{\boldsymbol{x}^{(i)}}\right\}_{i=1}^{n_0}$ is generated using Hammersley sequence within the feasible truncated domain $D_{\delta_0}$.
$D_{\delta_0}$ is similar with $D_{\delta_1}$ by replacing $\delta_0$ with $\delta_1$. 
Then, the log-likelihood function $\mathscr{L}(\boldsymbol{x})$ is evaluated at these points to produce the corresponding response values, i.e., $\boldsymbol{\mathfrak{L}} =\{\mathscr{L}(\boldsymbol{x}^{(i)})\}_{i=1}^{n_0}$.
Finally, the initial training set is constructed by $\boldsymbol{\mathcal{D}}=\{\boldsymbol{\mathcal{X}},\boldsymbol{\mathfrak{L}}\}$.
Let $n = n_0$. 

\textbf{Step 4: Build a GPR model}

Train a GP model $\mathscr{L}_{n}(\omega,\boldsymbol{x})$ to approximate the log-likelihood function using the training dataset $\boldsymbol{\mathcal{D}}=\{\boldsymbol{\mathcal{X}},\boldsymbol{\mathfrak{L}}\}$.
In the present study, the fitrgp function from Matlab R2024a with a constant prior mean and an squared exponential kernel (see Eq.~\eqref{eq:prior_covariance} is employed.

\textbf{Step 5: Estimate the model evidence}

Estimate the plug-in model evidence estimator $\hat{\tilde{c}}_n$ using MCS over $\mathbb{S}$, as well as the lower bound $\hat{\underline{c}}_n$ and the upper bound $\hat{\overline{c}}_n$.

\textbf{Step 6: Check the stopping criterion \#1}

If $\frac{\hat{\overline{c}}_n-\hat{\underline{c}}_n}{\hat{\tilde{c}}_n} < \epsilon_{\text{RE}}$ is satisfied for two successive iterations, go to \textbf{Step 8}; Otherwise, go to \textbf{Step 7}. 

\textbf{Step 7: Enrich the training dataset}

Identify the next point by solving 
$\boldsymbol{x}^{(n+1)} 
= \underset{\boldsymbol{x} \in D_{\delta_1}}{\arg\max} \, \mathcal{LF}(\boldsymbol{x})$ using any appropriate optimization algorithm such as, e.g. starfish optimization algorithm (SFOA)~\cite{zhong2025starfish}.
Evaluate the log-likelihood function at $\boldsymbol{x}^{(n+1)}$ to obtain $\boldsymbol{\mathfrak{L}}^{(n+1)} = \mathscr{L}(\boldsymbol{x}^{(n+1)})$, and update the training set: $\boldsymbol{\mathcal{D}}= \boldsymbol{\mathcal{D}}\cup\{\boldsymbol{x}^{(n+1)},\boldsymbol{\mathfrak{L}}^{(n+1)}\}$.
Set $n=n+1$ and return to \textbf{Step 4}.

\textbf{Step 8: Check the stopping criterion \#2}

Calculate the CoV of plug-in model evidence estimator $\hat{\tilde{c}}_n$, denoted as $\mathrm{CoV}[\, \hat{\tilde{c}}_n]$.
If $\mathrm{CoV}[\,\hat{\tilde{c}}_n] \leq \eta$, proceed to \textbf{Step 10}; otherwise, go to \textbf{Step 9}.

\textbf{Step 9: Enrich the sample pool}

To reduce the sampling variability in the plug-in model evidence estimator, enlarge the current sample set. 
Generate an additional set of $N$ samples, denoted as $\mathbb{S}^+$, from the prior distribution as described in \textbf{Step 2}.
Update the existing sample pool with the new sample, i.e., $\mathbb{S}=\mathbb{S}\cup{\mathbb{S}^+}$, and proceed to \textbf{Step 5}.

\textbf{Step 10: Approximate the posterior PDF}

Once the plug-in model evidence estimator $\hat{\tilde{c}}_n$ and the posterior mean function $e^{m_{\mathscr{L}_n}(\boldsymbol{x})}$ for the likelihood function have been adequately estimated, the posterior PDF can be constructed. 
The approximation of the posterior PDF is given by:
\begin{equation} 
\hat{f}(\boldsymbol{x}|\boldsymbol{Y}_{\mathrm{obs}})=\frac{e^{m_{\mathscr{L}_n}(\boldsymbol{x})}f(\boldsymbol{x})}{\hat{\tilde{c}}_n}.
\label{eq:posterior_estimate}
\end{equation}

With the posterior PDF available, the mean and standard deviation of the model parameters can be directly evaluated via numerical integration; in this study, the MCS is adopted.
Furthermore, if posterior samples are of interest, they can be generated using sampling strategies based on the posterior PDF, sampling-importance-resampling (SIR)~\cite{smith1992bayesian}, sequential importance sampling, or acceptance-rejection sampling. 
In this study, the SIR method is used.

\section{Numerical examples}
\label{sec:Numerical examples}
In this section, four numerical examples are presented to illustrate the performance of the proposed method.
In each of these examples, the parameters in the proposed SALBC method are set as: $n_0 = 4\sim10$, $N_M = 2\times10^4$, $N = 2\times10^4$, $b = 1$, $\epsilon_{\text{RE}} = 0.1$, $\eta = 0.02$, $\delta_0 =1 \times 10^{-2}$, and $\delta_1 = 1\times 10^{-5}$.  
For comparison, several state-of-the-art methods, including TMCMC~\cite{ching2007transitional, betz2016transitional}, BUS~\cite{Straub2015,betz2018bayesian}, and ABQ-BU~\cite{song2025sampling} are evaluated.
To ensure the robustness of the results, both the proposed method and the comparison methods are repeated 20 times, and the corresponding CoV are calculated. 
For the last two examples, however, the TMCMC and BUS are only conducted once due to infeasibility.
All simulations are performed on a Windows 10 system equipped with an Intel(R) Core(TM) i9-14900K processor (2.19 GHz), 60 GB RAM, and a 64-bit architecture.

\subsection{Example 1: An illustrative example}
The first example considers a sigmoid function, which has an analytical solution for illustrative purposes.
This example involves a single random variable $X$, with a prior distribution following the Gaussian distribution $N$(1.5, $2^2$).
The corresponding response function is expressed as follows:
\begin{equation} 
\mathcal{R}(X)=\frac{10}{1 + e^{-1.2(X-1)}}.
\label{Eq1}
\end{equation}

In this example, the observation data is set to $\mathcal{R}(1)$ = 5. 
To construct the likelihood function, we assumed that the prediction error follows a zero-mean Gaussian distribution with a standard deviation of $\sigma_\epsilon$ = 0.2.
According to Eq.~\eqref{eq:likelihood_explicit}, the likelihood function is expressed as:
\begin{equation} 
L(Y_{\mathrm{obs}}|x))=\exp\left( -\frac{(\mathcal{R}(1)-\mathcal{R}(x))^2}{2\times0.2^2}\right).
\label{eq:likelihood_ex1}
\end{equation}

Since the likelihood function is a deterministic function of the random variable $X$, it enables a semi-analytical solution for the posterior PDF, which serves as a reference for comparison.

\begin{figure}[htbp]
    \centering
    \subfigure[GPR model $n=n_0 = 4$]{
        \includegraphics[scale=0.25]{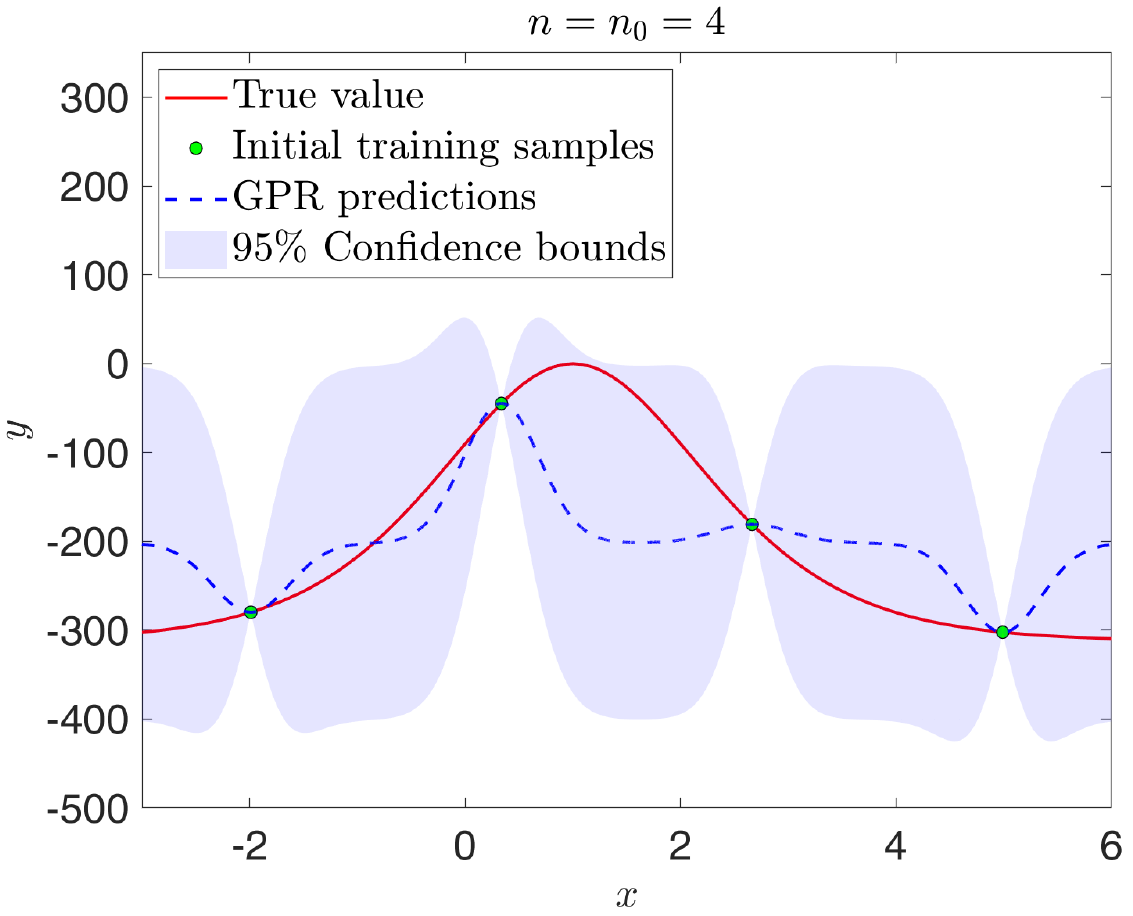} \label{fig:enter-label1}
    }
    \subfigure[GPR model $n = 5$]{
        \includegraphics[scale=0.25]{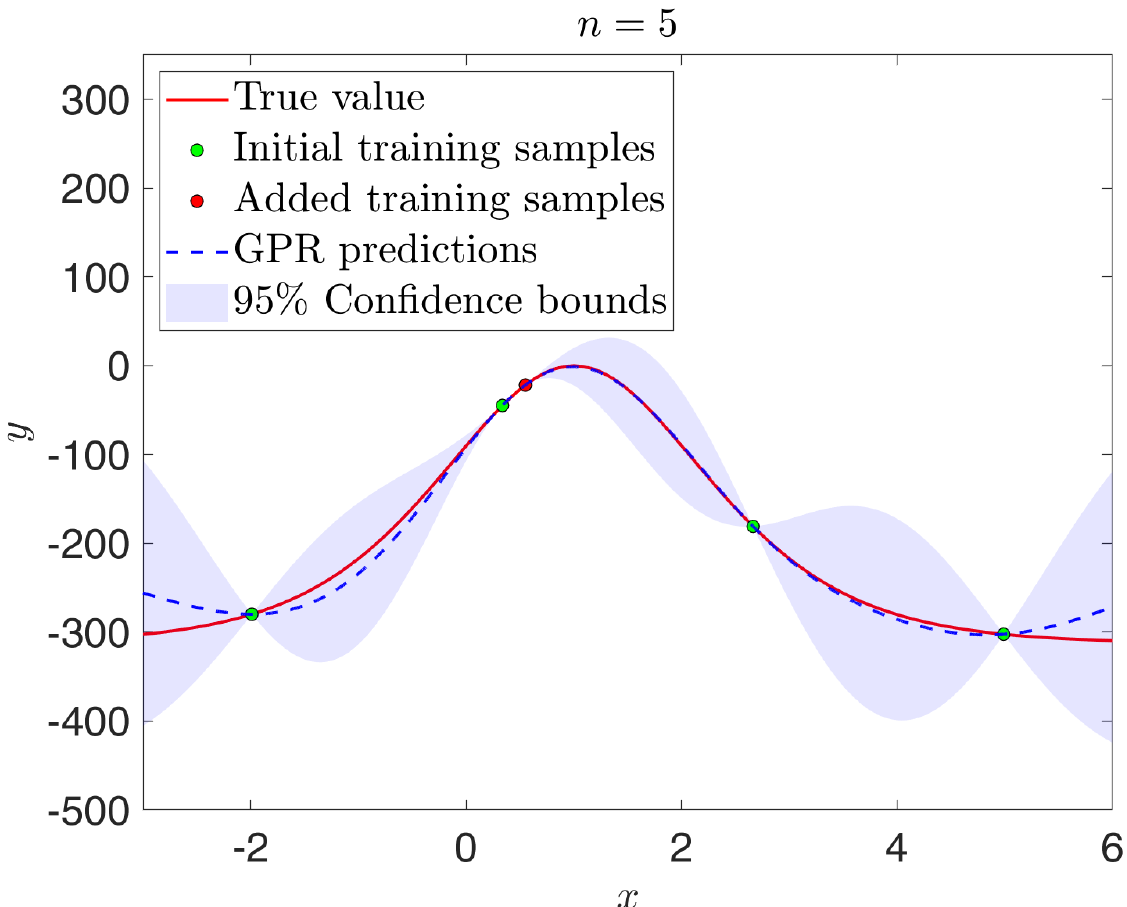} \label{fig:enter-label2}
    }
    \subfigure[GPR model $n = 8$]{
        \includegraphics[scale=0.25]{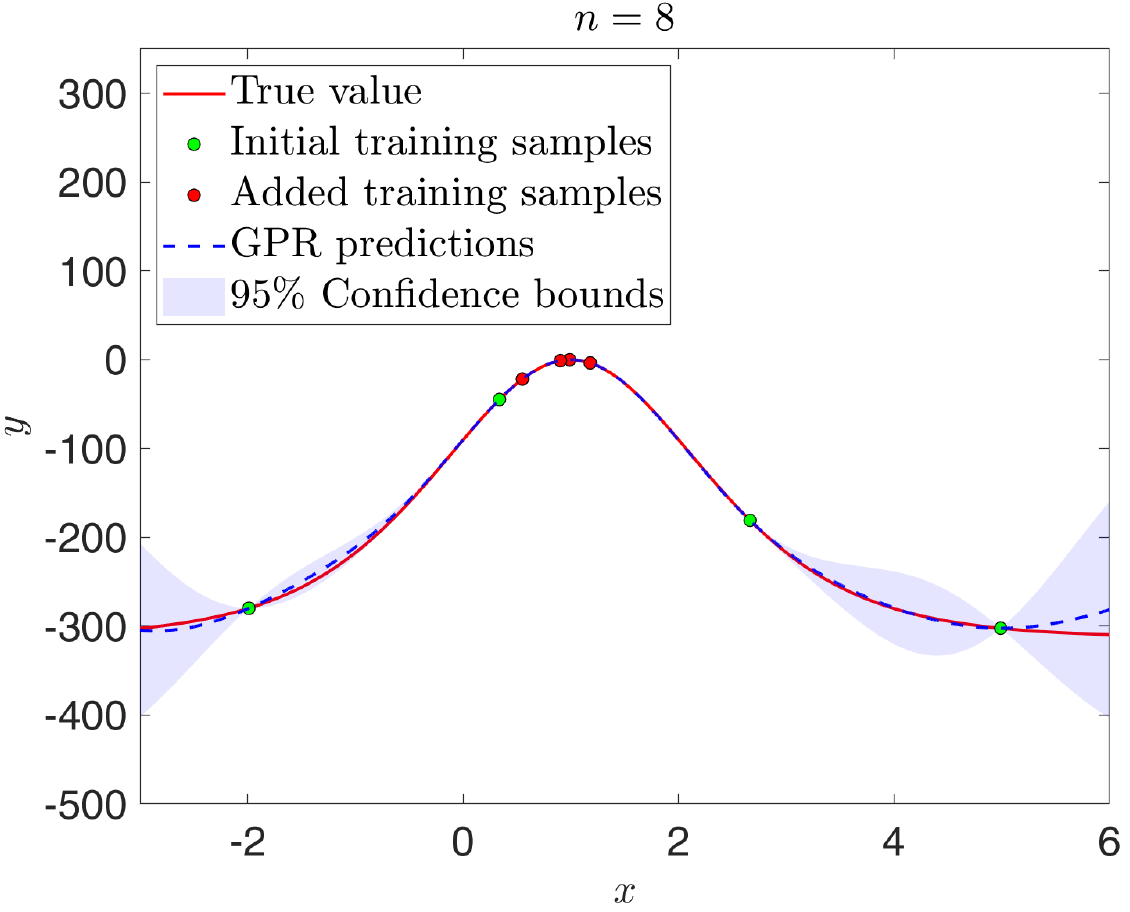} \label{fig:enter-label3}
    }
    \vspace{0.5cm} 

    \subfigure[Learining function $n=n_0 = 4$]{
        \includegraphics[scale=0.25]{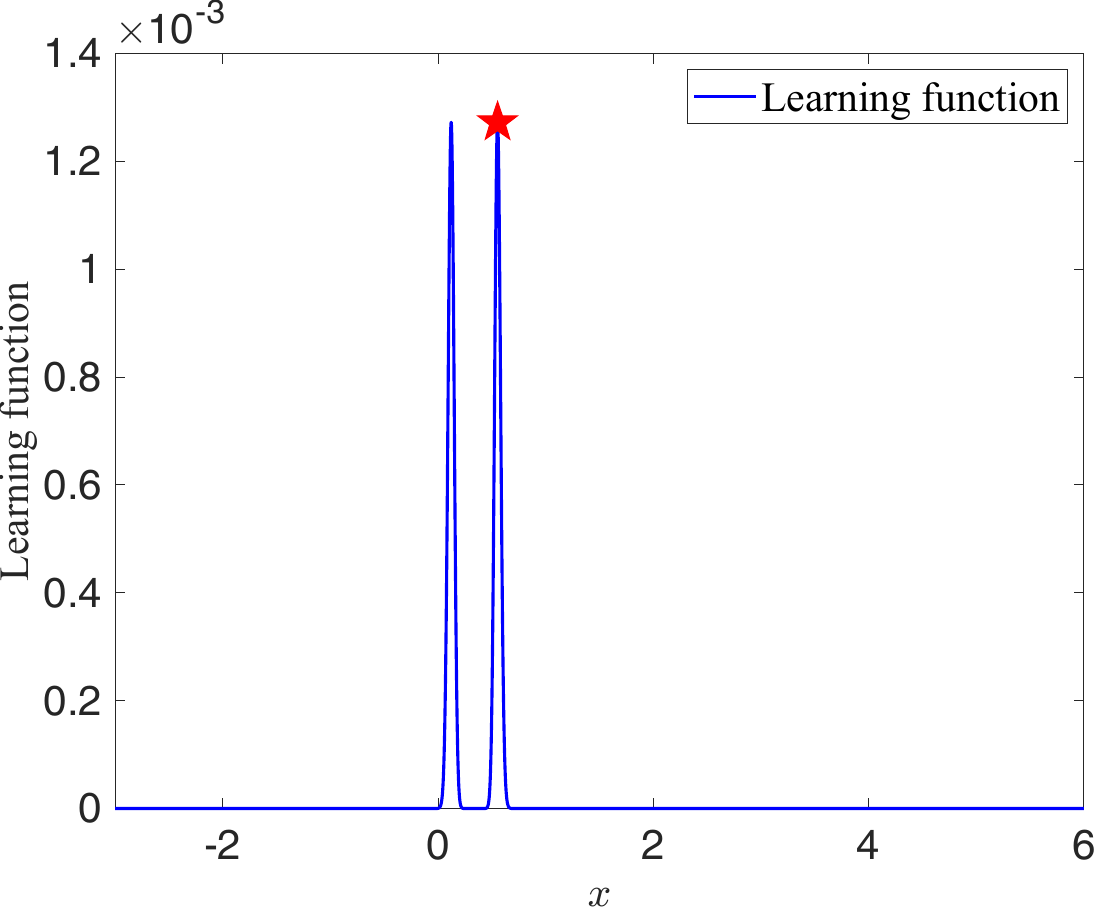} \label{fig:enter-label4}
    }
    \subfigure[Learining function $n= 5$]{
        \includegraphics[scale=0.25]{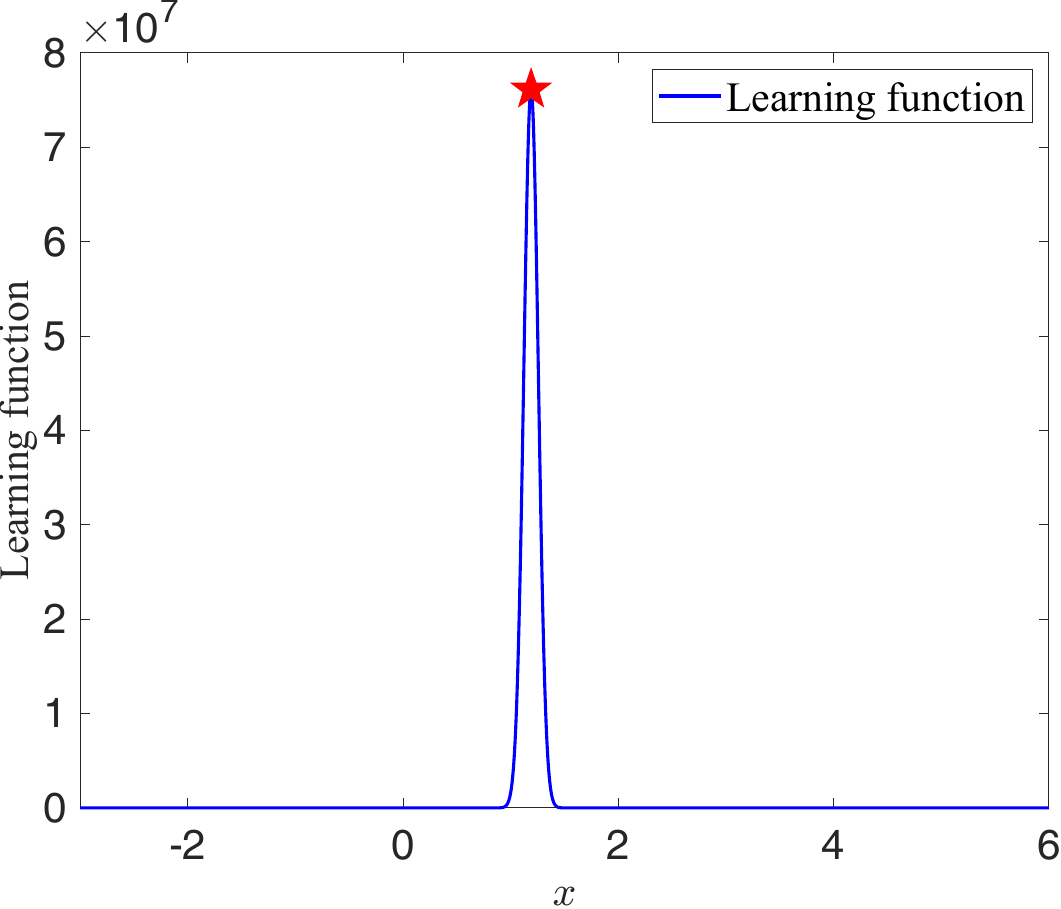} \label{fig:enter-label5}
    }
    \subfigure[Learining function $n = 8$]{
        \includegraphics[scale=0.25]{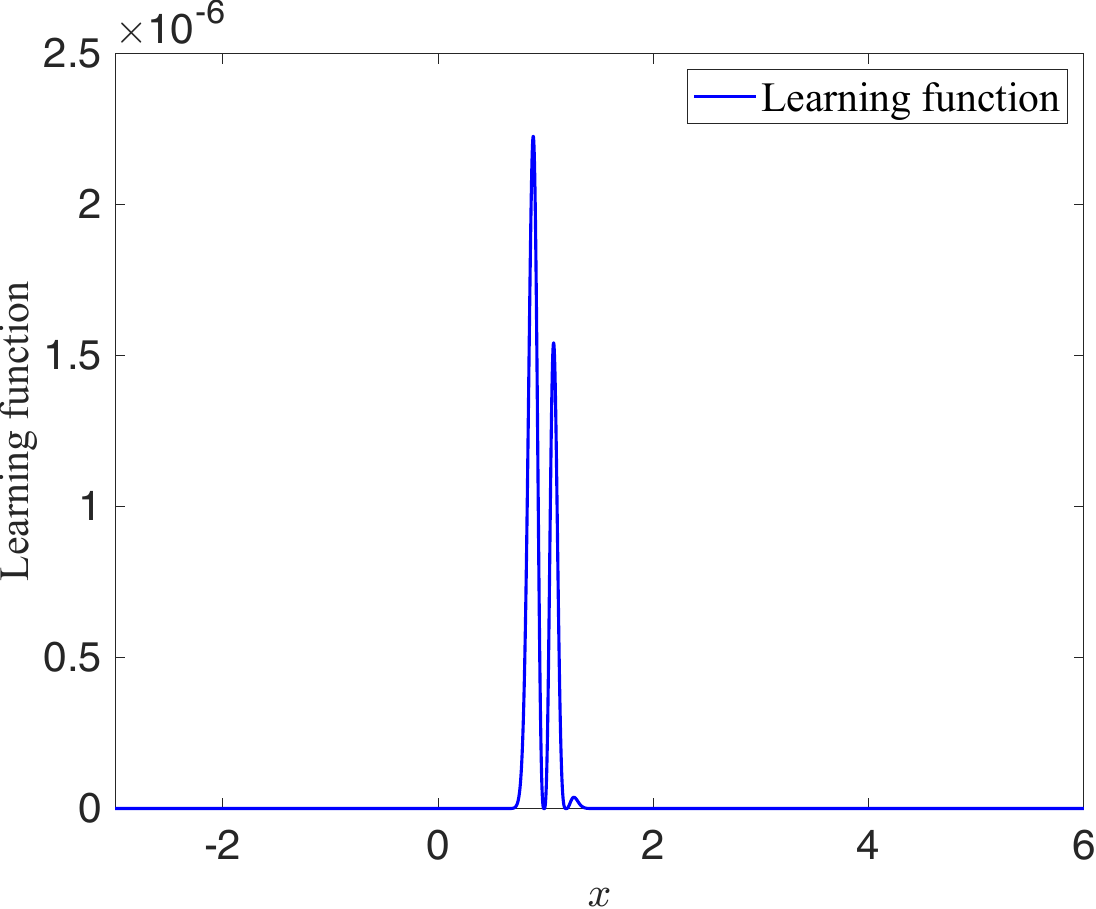} \label{fig:enter-label6}
    }
    \vspace{0.5cm} 

    \subfigure[Posterior PDF $n=n_0 = 4$]{
        \includegraphics[scale=0.25]{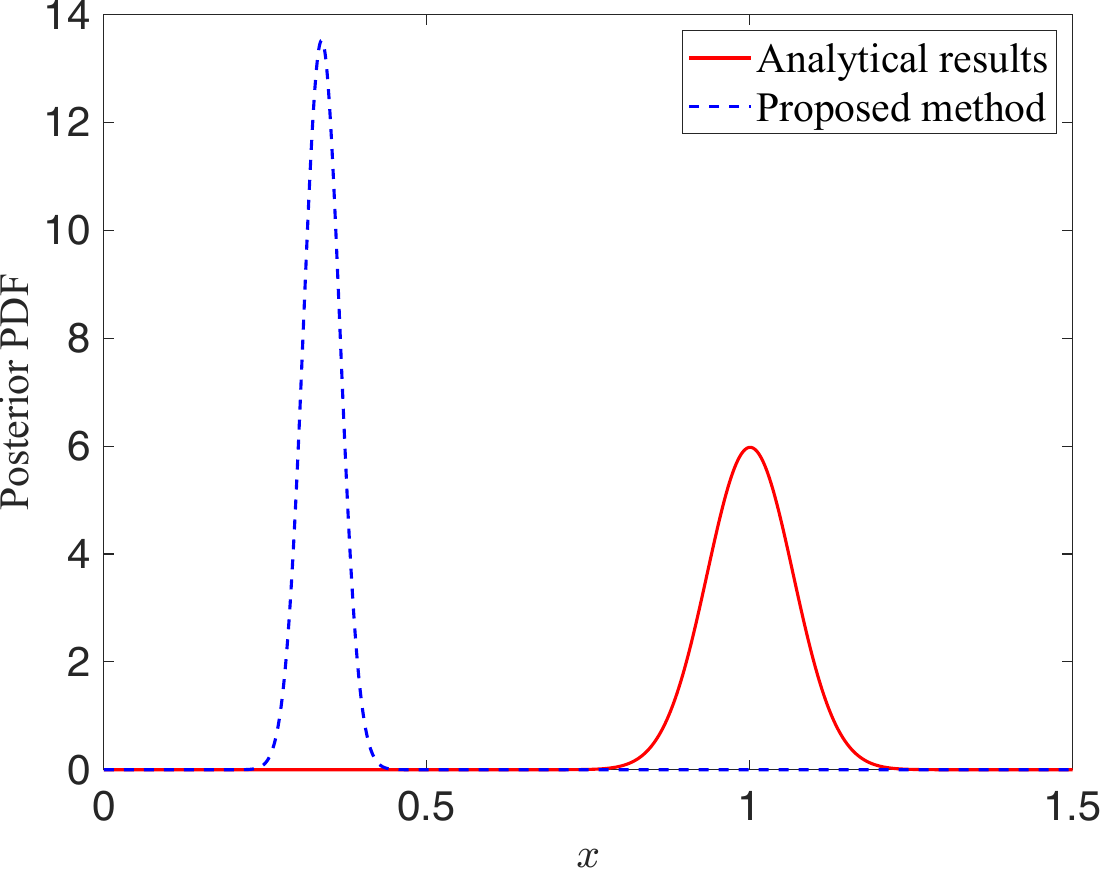} \label{fig:enter-label7}
    }
    \subfigure[Posterior PDF $n = 5$]{
        \includegraphics[scale=0.25]{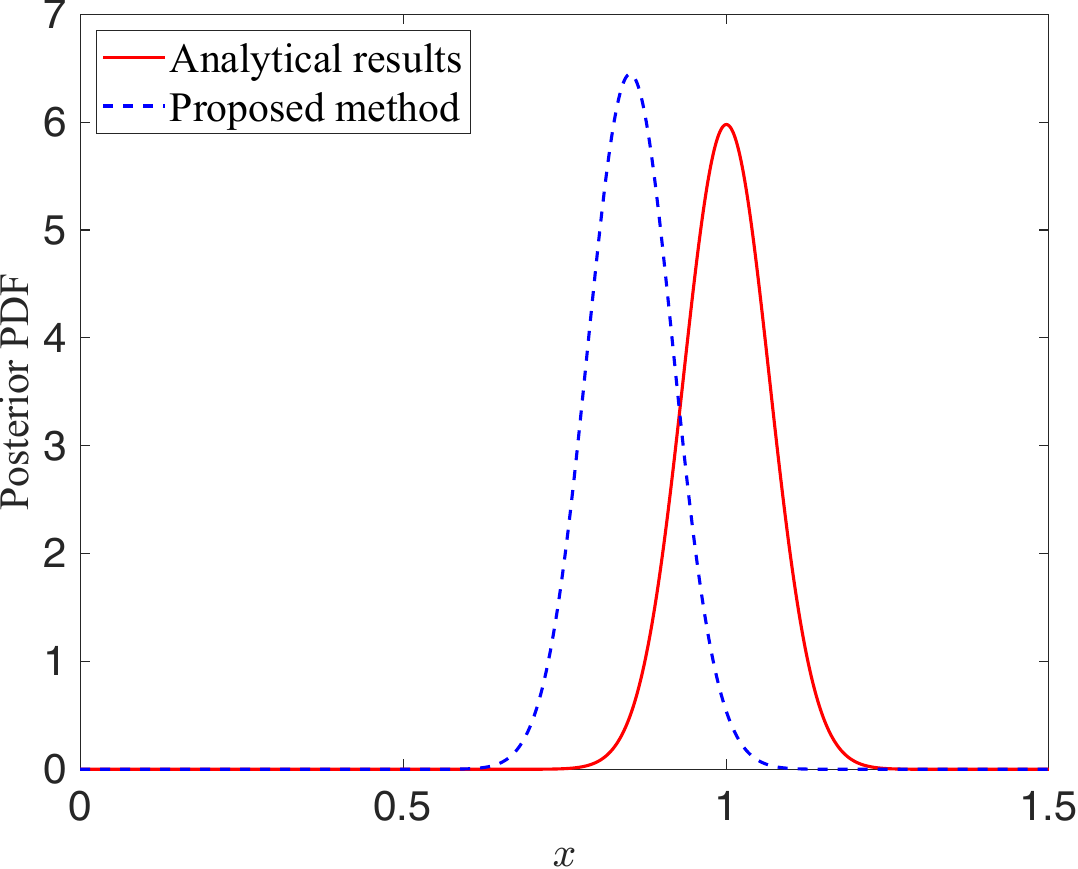} \label{fig:enter-label8}
    }
    \subfigure[Posterior PDF $n = 8$]{
        \includegraphics[scale=0.25]{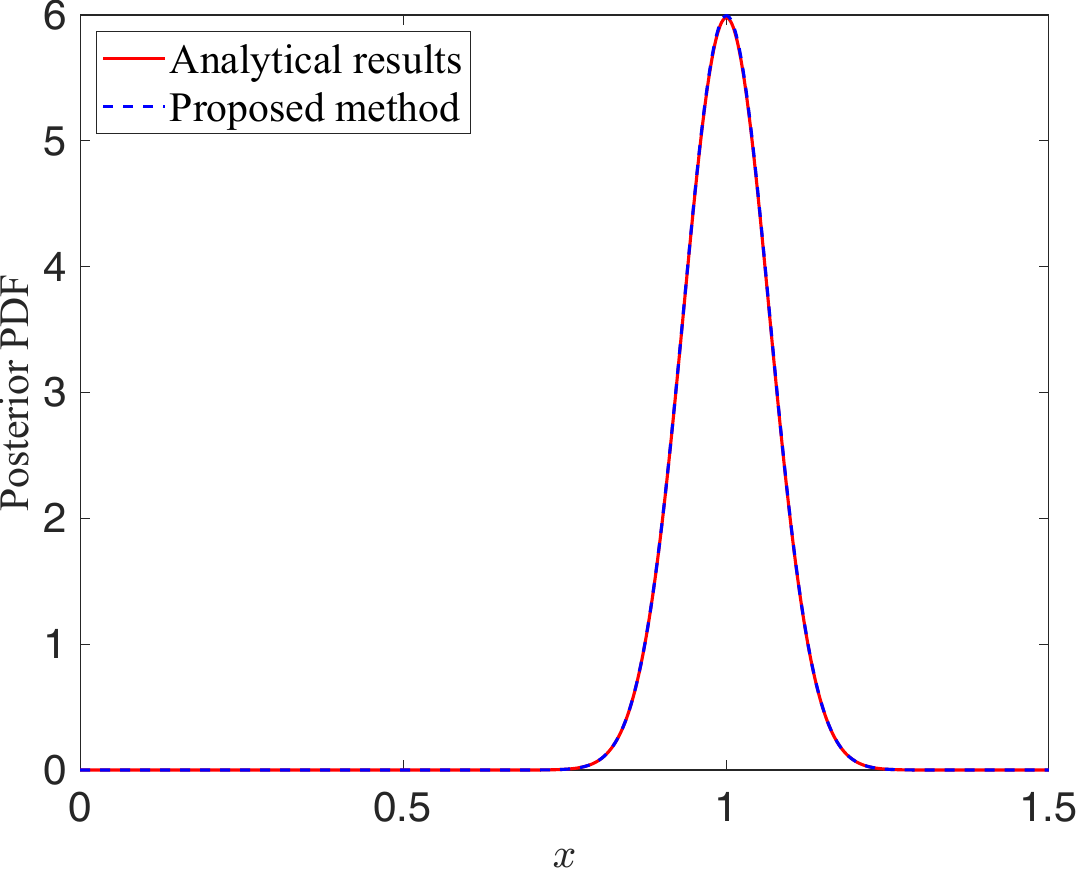} \label{fig:enter-label9}
    }
    \caption{Illustration of the computational details for Example 1}
    \label{fig:details_ex1}
\end{figure}

Fig.~\ref{fig:details_ex1} shows the computational details of the proposed method with an initial sample size of $n_0 = 4$.
Each column corresponds to a different number of training samples ($n$ = 4, 5, 8), and includes three subplots: (1) the GPR model plot with $95\%$ prediction interval, (2) the learning function $\mathcal{LF}(\boldsymbol{x})$ with the optimal next training point, and (3) the approximate posterior PDF compared to the analytical solution.
At $n=n_0$ = 4, where only the initial training points is used, the uncertainty of GPR model is large, as indicated by the wide prediction interval, and the estimated posterior PDF significantly deviates from the analytical result. 
The learning function effectively captures the most informative region in the GPR model and guides the selection of an optimal next training point.
When one new point is added ($n = 5$), the uncertainty is greatly reduced, leading to a posterior PDF that more closely approximates the analytical one. 
By $n = 8$, the prediction interval becomes narrow, and the estimated posterior PDF aligns very well with the analytical solution, demonstrating the efficiency and effectiveness of the proposed active learning method.

\begin{table}[h]
\centering
\renewcommand{\arraystretch}{1.2} 
\caption{Model updating results for Example 1}\label{tab:reults_ex_1}
\begin{tabular}{ll@{\hspace{20pt}}lllllllll}
\hline
      \multirow{2}{*}{Method} & \multirow{2}{*}{Time(s)} & \multicolumn{2}{c}{$N_{call}$ }    &  \multicolumn{2}{c}{$\hat{c}$}   &  \multicolumn{2}{c}{$\hat{\mu}_{X}$}     &  \multicolumn{2}{c}{$\hat{\sigma}_{X}$}  \\ 
      \cmidrule(r){3-4}   \cmidrule(r){5-6} \cmidrule(r){7-8}  \cmidrule(r){9-10} 
      & & Mean & CoV & Mean & CoV & Mean & CoV & Mean & CoV \\
\hline                             
Reference        &         &         &         & 0.0323  & 0.00\%  & 1.0006 & 0.00\%  & 0.0668 &  0.00\% \\ 
BUS              & 11.42        &  10000       & 0.00\%        & 0.0324        & 4.45\%         &1.0010         &  0.87\%        &  0.0667      &   1.19\%    \\
TMCMC            &  95.20       &  10000   &  0.00\%  & 0.0322        &   2.26\%          &   1.0002      &   0.83\%      &  0.0665      &       0.91\%       \\
ABQ-BU           &   636.24      &    8.6   &  6.00\%  &    0.0317      &   4.71\%      &   0.9998       &   0.44\%      &  0.0694      &   7.79\%      \\
Proposed method  &     21.34 &     8.0     & 0.00\%  &  0.0323     &     1.55\%      &     1.0001     &        0.06\%  &    0.0669    &       0.72\%       \\                             
\hline
\end{tabular}
\label{tab:reults_ex_1}
\end{table}

To further assess the performance of the proposed method, comparisons are made with BUS, TMCMC, and ABQ-BU. 
The comparison results, including computational time, the number of model evaluations ($N_{call}$), model evidence estimation ($\hat{c}$), and the posterior mean ($\hat{\mu}_{X}$) and standard deviation ($\hat{\sigma}_{X}$) of $X$, are presented in Table~\ref{tab:reults_ex_1}.
As shown in Table~\ref{tab:reults_ex_1}, the model evidence estimation, posterior mean and standard deviation of $X$ obtained from the proposed method closely match the reference values of the analytical PDF, while maintaining a low CoV that indicates high robustness. 
Moreover, compared to BUS, TMCMC, and ABQ-BU, the proposed method requires fewer model evaluations and attains the second-lowest computational time. 

\subsection{Example 2: Three degree-of-freedom (DOF) mass-spring system}
\begin{figure}
    \centering
    \includegraphics[width=0.6\linewidth]{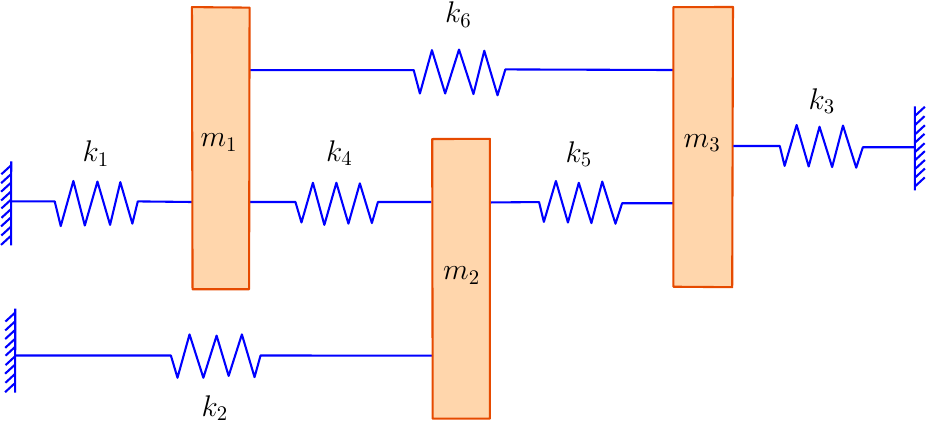}
    \caption{Three degree-of-freedom spring mass system}
    \label{Ex2_three_freedom_model}
\end{figure}

The second example considers a three DOF mass-spring system, as shown in Fig.~\ref{Ex2_three_freedom_model}. 
This system has been widely investigated in previous studies~\cite{bi2019role,mares2006stochastic,khodaparast2008perturbation,patelli2017sensitivity}. 
In this system, the nominal mass values are set to $m_1=m_2=m_3=1.0$~kg, and the nominal stiffness values are specified as $k_1 = k_2 = k_5$ = 1.5 N/m, $k_3 = k_4$ = 1.0 N/m, and $k_6$ = 3.0 N/m.
In this study, the stiffness parameters $k_{1}$, $k_{2}$, and $k_{5}$ are treated as uncertain variables to be updated.
Their prior distributions are assumed as Gaussian distribution with mean values $\mu_{k_{1}}$ = $\mu_{k_{2}}$ = $\mu_{k_{5}}$= 2.0 N/m and standard deviations $\sigma_{k_{1}}=\sigma_{k_{2}}=\sigma_{k_{5}}=$0.3 N/m.
The uncertainty in $k_1$, $k_2$, and $k_5$ propagates through the system and influences the observed natural frequencies $f_{\mathrm{obs},1}, f_{\mathrm{obs},2}, f_{\mathrm{obs},3}$.

A total of 30 sets of observations are generated using nominal mass values and stiffness parameters $k_1$, $k_2$, and $k_5$, where the means are set to their nominal stiffness values and the standard deviations to $\sigma_{k_1} = \sigma_{k_2} = \sigma_{k_5} = 0.2$ N/m.
With the above setting, the likelihood function can be expressed as:
\begin{equation}
L(\boldsymbol{Y}_{\mathrm{obs}}\mid \boldsymbol{x})
= \prod_{r=1}^{30} \exp\!\left(
    -\frac{1}{2\sigma_\epsilon^2}
    \sum_{i=1}^3 \big(f_{\mathrm{obs},i}^{(r)} - f_i(\boldsymbol{x}) \big)^2
  \right)
= \exp\!\left(
    -\frac{1}{2\sigma_\epsilon^2}
    \sum_{r=1}^{30}\sum_{i=1}^3 \big(f_{\mathrm{obs},i}^{(r)} - f_i(\boldsymbol{x})\big)^2
  \right),
\end{equation}
where $\sigma_\epsilon = 0.01$, $f_{\mathrm{obs},i}^{(r)}$ denotes the $r$-th observation set corresponding to the $i$-th natural frequency, and $f_i(\boldsymbol{x})$ denotes the $i$-th natural frequency, obtained as the system response for a given realization of $k_1$, $k_2$, and $k_5$.

To evaluate the performance of the proposed method, commonly used updating techniques, including BUS, TMCMC, and ABQ-BU, are employed to estimate the posterior distribution of $k_1, k_2, k_5$.
The comparative results, including computational time, the number of model evaluations, the estimated model evidence $\hat{c}$, and the posterior mean and standard deviation of $k_1, k_2, k_5$ are summarized in Table~\ref{tab:reults_ex_2}. 
From Table~\ref{tab:reults_ex_2}, it can be seen that the results obtained using the proposed method ($n_0=4$) are close to those obtained from the other methods and exhibit a low CoV, indicating the robustness of the proposed method. 
Notably, the proposed method requires significantly fewer model evaluations (only 23.3) than the other methods and achieves a much shorter computational time (9.65 s), second only to BUS. 

\begin{table}[h]
\small
\centering
\renewcommand{\arraystretch}{1.2} 
\caption{Model updating results for Example 2}\label{tab:reults_ex_2}
\begin{tabular}{l@{\hspace{30pt}}llllllllllll}
\hline
Method & BUS & TMCMC & ABQ-BU & Proposed method  \\
\hline
{Time(s)} & 4.92 & 37.06 &  27970.15 & 9.65 \\
{$N_{call}$ (CoV)} & 10000  & 10000 & 144.15 (43.66\%) & 23.3 (3.71\%) \\
{$\hat{c}$ (CoV)}  & $3.1478 \times 10^{-14}$  (11.22\%)& $3.0598 \times 10^{-14}$  (10.95\%)  & $3.2542 \times 10^{-14}$  (46.2\%) & $3.2107 \times 10^{-14}$ (9.02\%)  \\
{$\hat{\mu}_{k_1}$ (CoV)}  & 1.5342 (0.30\%) & 1.5333 (0.15\%)   & 1.5505 (2.51\%) & 1.5331 (0.39\%)  \\
{$\hat{\mu}_{k_2}$ (CoV)}  & 1.4459 (0.39\%)  & 1.4463 (0.19\%)   & 1.4296 (2.62\%) & 1.4550 (0.46\%)  \\
{$\hat{\mu}_{k_5}$ (CoV)}  & 1.4439 (0.15\%) & 1.4441 (1.11\%)  &  1.4467 (1.21\%) & 1.4550 (0.19\%)  \\
{$\hat{\sigma}_{k_1}$ (CoV)} & 0.0989 (2.52\%) & 0.0988 (1.24\%) &  0.0816 (29.38\%) & 0.0970 (2.77\%)   \\
{$\hat{\sigma}_{k_2}$ (CoV)} & 0.1049 (2.33\%) & 0.1035 (1.35\%) &  0.0890 (28.26\%) & 0.1027 (2.77\%)  \\
{$\hat{\sigma}_{k_5}$ (CoV)} & 0.0501 (1.91\%) & 0.0496 (1.49\%) &  0.0416 (26.01\%) & 0.0501 (2.54\%)  \\
\hline
\end{tabular}
\label{tab:reults_ex_2}
\end{table}

\begin{figure}[htbp]
    \centering
    \subfigure[$f_1$ vs $f_2$]{
        \includegraphics[scale=0.25]{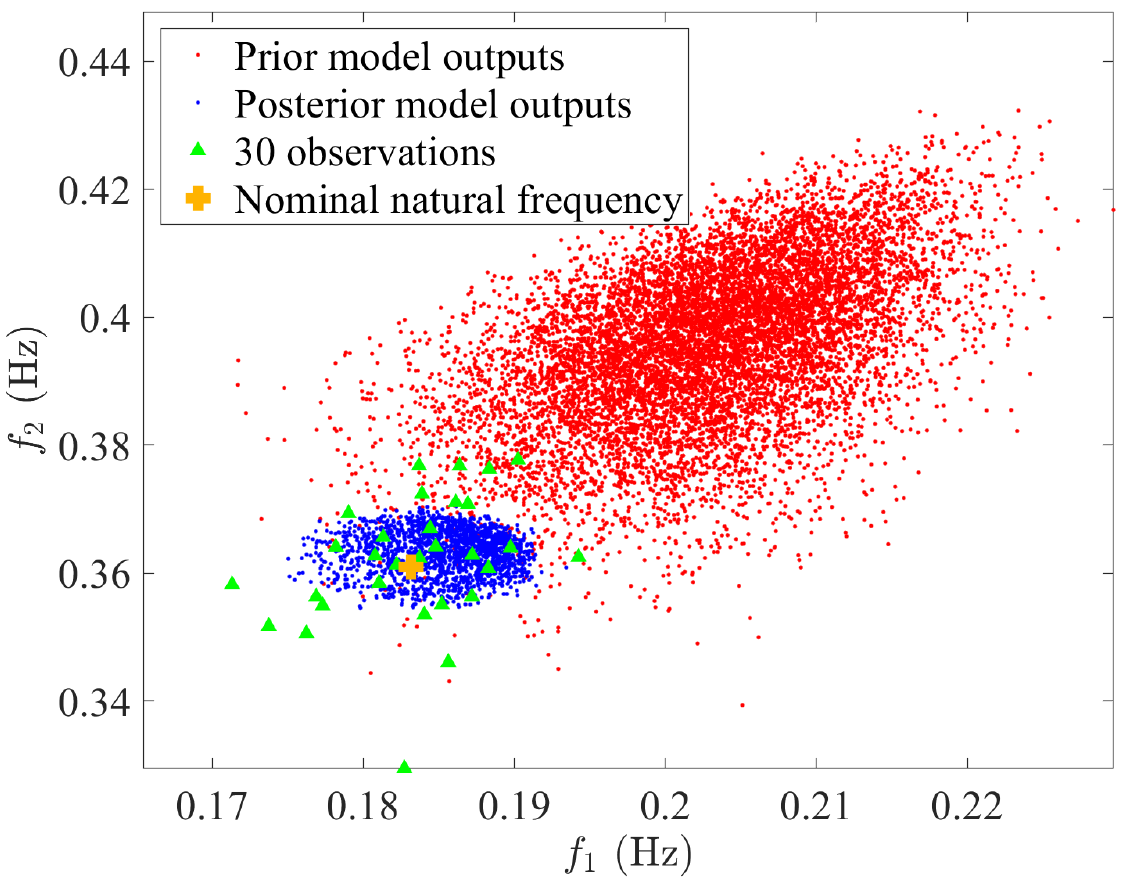} \label{fig:enter-label1}
    }
    \subfigure[$f_2$ vs $f_3$]{
        \includegraphics[scale=0.25]{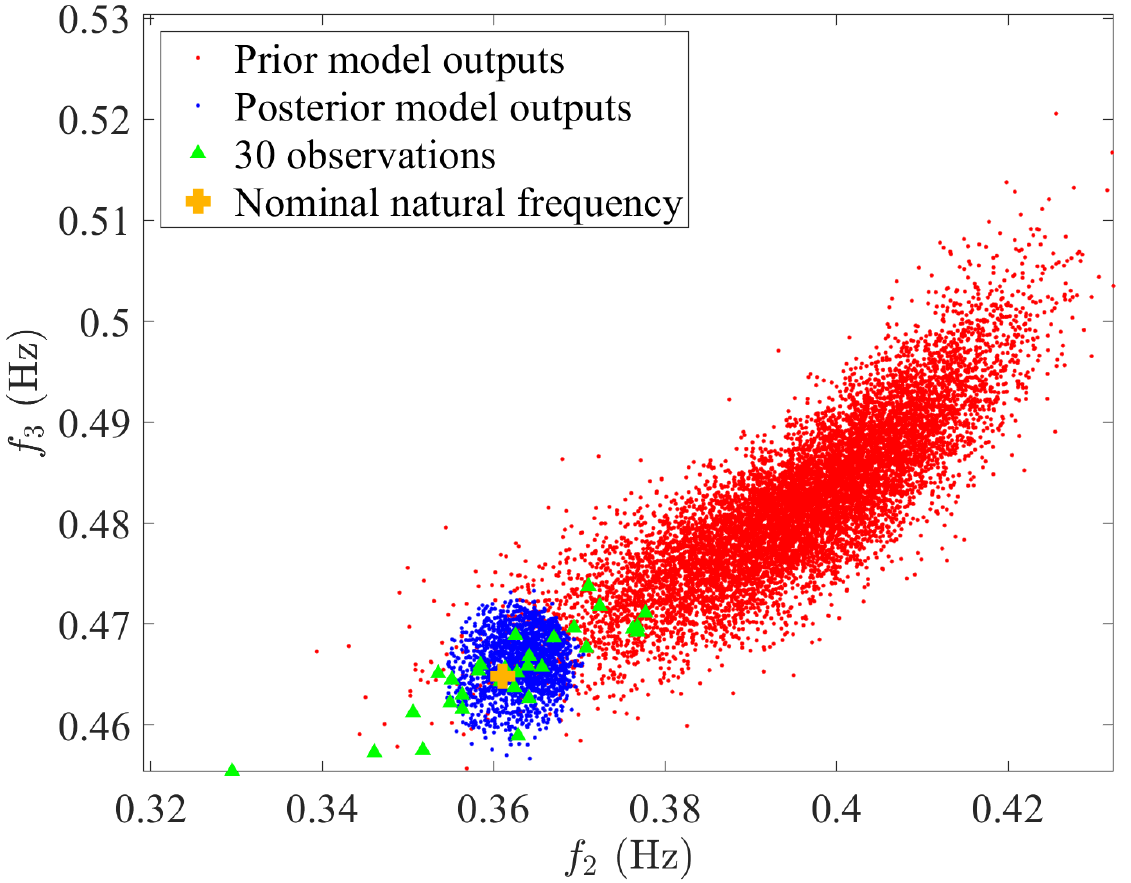} \label{fig:enter-label2}
    }
    \subfigure[$f_1$ vs $f_3$]{
        \includegraphics[scale=0.25]{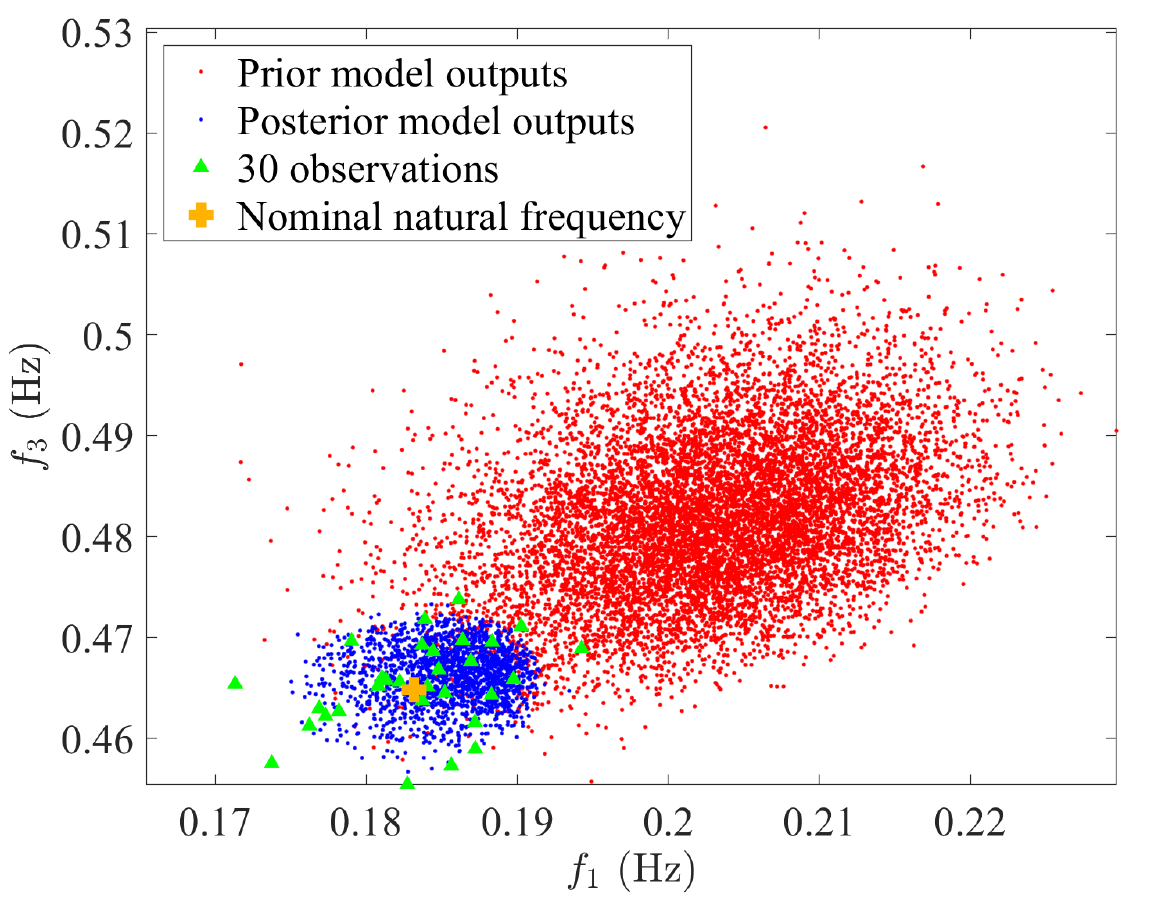} \label{fig:enter-label3}
    }
    \caption{The observations and the output samples before and after updating for Example 2}
\label{fig:ex2_samples_output}
\end{figure}

After determining the posterior distributions of $k_1, k_2, k_5$, the corresponding samples of the model outputs $f_1, f_2, f_3$ can be easily obtained. 
Fig.~\ref{fig:ex2_samples_output} illustrates the initial model outputs, the model outputs after updating using the proposed method, and the observed values.
As shown in Fig.~\ref{fig:ex2_samples_output}, the outputs of the prior model (in red) display a significantly wider spread and greater uncertainty in all frequency pair spaces compared to the posterior samples (in blue) obtained using the proposed method. 
After updating, the posterior samples are clustered around the observed natural frequencies (green triangles) and align closely with the nominal value obtained from the corresponding nominal values of the stiffness (orange cross), demonstrating the effectiveness of the proposed method.

\subsection{Example 3: Modal analysis for the shoulder link of a lightweight robotic arm}
\begin{figure}
    \centering
    \includegraphics[width=0.7\linewidth]{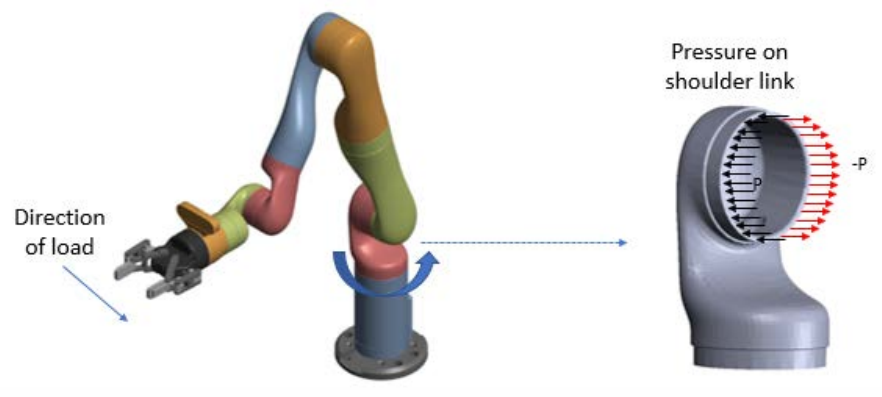}
    \caption{The shoulder link of a robotic arm}
    \label{fig:ex3_model_geometry}
\end{figure}

\begin{figure}[htbp]
    \centering
    \subfigure[FEM]{
        \includegraphics[scale=0.32]{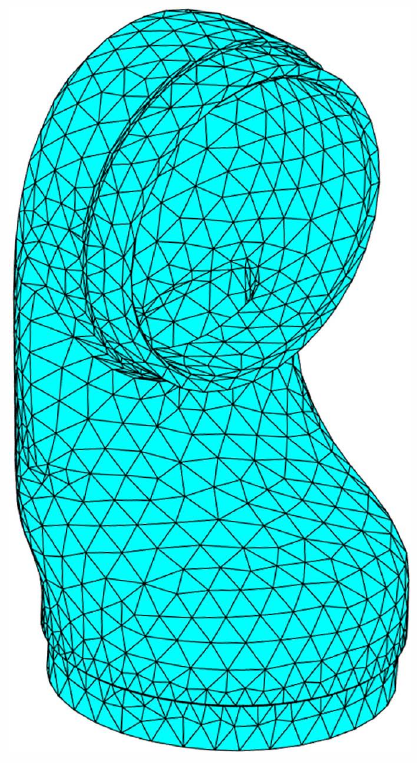} \label{fig:enter-label1}
    }
    \subfigure[Mode 1]{
        \includegraphics[scale=0.7]{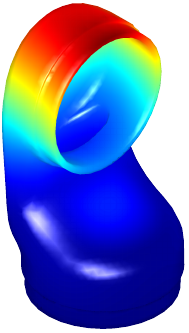} \label{fig:enter-label2}
    }
    \subfigure[Mode 2]{
        \includegraphics[scale=0.7]{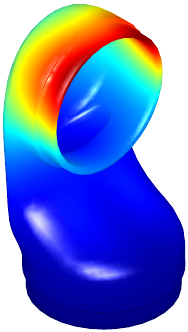} \label{fig:enter-label3}
    }
    \vspace{0.5cm} 

    \subfigure[Mode 3]{
        \includegraphics[scale=0.7]{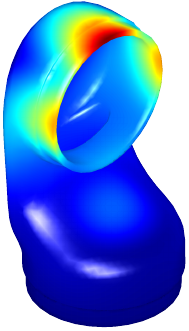} \label{fig:enter-label4}
    }
    \subfigure[Mode 4]{
        \includegraphics[scale=0.7]{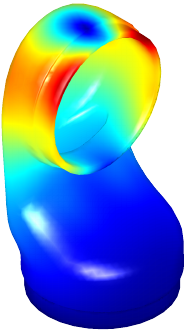} \label{fig:enter-label5}
    }
    \subfigure[Mode 5]{
        \includegraphics[scale=0.7]{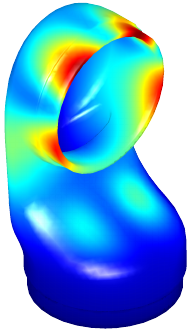} \label{fig:enter-label6}
    }
    
    \caption{The finite element mesh and first five modes of the robotic arm's shoulder link.}
\label{fig:ex3_mesh_modes}
\end{figure}

The third example involves the shoulder link of a lightweight robotic arm, was previously investigated by Song et al.~\cite{song2025sampling}.
The configuration of this shoulder link is illustrated in Fig.~\ref{fig:ex3_model_geometry}.
The material of the shoulder link is assumed to be homogeneous, three material properties including Young’s modulus ($E$), Poisson’s ratio ($\nu$), and mass density($\rho$) are considered as uncertain parameters.
These uncertain parameters influence the natural frequencies and mode shapes of the structure and are subsequently calibrated using Bayesian updating.
The model analysis is performed using the PDE toolbox in Matlab.
The corresponding finite element mesh and the first five modes of the robotic arm's shoulder link (for nominal values of $E$, $\nu4$ and $\rho$) are shown in Fig.~\ref{fig:ex3_mesh_modes}.

The prior distribution of uncertain parameters is summarized in Table~\ref{tab:ex3_prior_information}. 
The first five natural frequencies were observed, namely $f_{\mathrm{obs},1}$ = 1947.2 Hz, $f_{\mathrm{obs},2}$ = 2662.0 Hz, $f_{\mathrm{obs},3}$ = 4982.3 Hz, $f_{\mathrm{obs},4}$ = 5112.6 Hz, and $f_{\mathrm{obs},5}$ = 7819.5 Hz, are used to update the distribution of $E$, $\nu$, and $\rho$.
Under this setup, the likelihood function is defined as:
\begin{equation}
L(\boldsymbol{Y}_{\mathrm{obs}}|\boldsymbol{x}) = \exp\left(-\frac{1}{2\sigma_\epsilon^2} \sum_{i=1}^{5} \left(\frac{f_{i}^2(\boldsymbol{x})}{f_{\mathrm{obs},i}^2} - 1 \right)^2 \right),
\label{eq:likelihood_ex3}
\end{equation}
where $\sigma_\epsilon = 0.2$, and $f_{i}(\boldsymbol{x})$ is the $i$-th natural frequency obtained from the model.

\begin{table}[h]
\centering
\renewcommand{\arraystretch}{1.2} 
\caption{Prior distribution of uncertain parameters for Example 3}\label{tab:information_ex_3}
\begin{tabular}{ccccc}
\hline
Parameter & Physical meaning & Distribution type & Lower bound & Upper bound \\
\midrule
$E$ & Young's modulus (GPa) & Uniform & 140 & 155 \\
$\nu$ & Poisson's ratio & Uniform & 0.28 & 0.38 \\
$\rho$ & Mass density (kg/m$^3$) & Uniform & 1900 & 2300 \\
\bottomrule
\end{tabular}
\label{tab:ex3_prior_information}
\end{table}

\begin{table}[h]
\centering
\renewcommand{\arraystretch}{1.2} 
\caption{Model updating results for Example 3}\label{tab:reults_ex_3}
\begin{tabular}{l@{\hspace{30pt}}llllllllllll}
\hline
Method & BUS & TMCMC & ABQ-BU & Proposed method  \\
\hline
{Time(s)} &   &  & 2,939.62 & 101.44 \\
{$N_{call}$ (CoV)} & 10000  & 10000 &  7.10 (4.34\%)  & 10 (0.00\%) \\
{$\hat{c}$ (CoV)}   & 0.6971 & 0.6358   & 0.6814 (6.44\%)  & 0.6860 (0.24\%) \\
{$\hat{\mu}_{E}$ (CoV)}  & $1.4811 \times 10^{11}$  &  $1.4824 \times 10^{11}$  & $1.4811 \times 10^{11}$ (0.01\%) & $1.4812 \times 10^{11}$ (0.03\%)  \\
{$\hat{\mu}_{\nu}$ (CoV)}  & 0.3298  & 0.3291      & 0.3293 (0.27\%)  & 0.3294 (0.10\%)  \\
{$\hat{\mu}_{\rho}$ (CoV)}   & $2.0605 \times 10^{3}$ & $2.0591 \times 10^{3}$  & $2.0639 \times 10^{3}$ (0.20\%) & $2.0627 \times 10^{3}$ (0.03\%)  \\
{$\hat{\sigma}_{E}$ (CoV)} & $4.2313 \times 10^{9}$  & $3.9543 \times 10^{9}$ & $4.2313 \times 10^{9}$ (0.79\%) & $4.270 \times 10^{9}$ (0.51\%)   \\
{$\hat{\sigma}_{\nu}$ (CoV)} & 0.0288  & 0.0271    & $0.0287$  (2.52\%) & $0.0289$  (0.43\%) \\
{$\hat{\sigma}_{\rho}$ (CoV)} & 106.0985 & 98.4536   & $105.1449$  (2.43\%)   &  $105.6342$  (0.55\%)  \\
\hline
\end{tabular}
\label{tab:reults_ex3}
\end{table}

Table~\ref{tab:reults_ex3} presents a comparison of the results obtained using BUS, TMCMC, ABQ-BU, and the proposed method.
Since this example involves computationally intensive finite element analysis (FEA), BUS and TMCMC were executed once, taking 3.21 and 2.67 days, respectively.
In contrast, both ABQ-BU and the proposed method were executed 20 times, with an initial sample size $n_0 = 5$. 
As shown in Table~\ref{tab:reults_ex3}, the estimated model evidence, the posterior mean and standard deviation of $E$, $\nu$, and $\rho$ obtained from the proposed method are consistent with those obtained from the other methods. 
Regarding computational efficiency, although the proposed method requires 10 model evaluations, slightly more than ABQ-BU (which averages 7.10 model evaluations), its total computational time is significantly lower (only 101.44 s).
This improvement is mainly attributed to the fact that the proposed method avoids the need to generate a large number of sample paths from the posterior GP.
Furthermore, the proposed method achieves a smaller CoV, which demonstrates its robustness.

\begin{figure}
    \centering
    \includegraphics[width=0.6\linewidth]{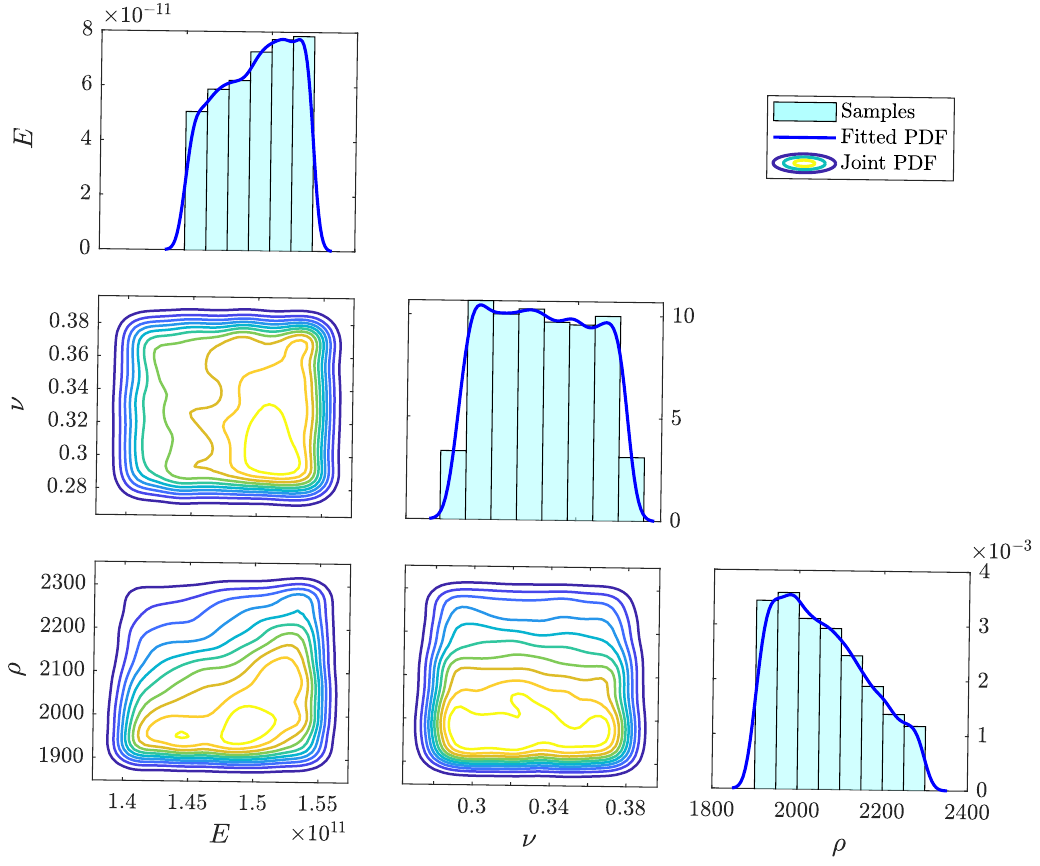}
    \caption{Posterior distributions of $E$, $\nu$, and $\rho$ in Example 3}
    \label{fig:Ex3_couter}
\end{figure}

Fig.~\ref{fig:Ex3_couter} shows the posterior PDFs of $E$, $\nu$, and $\rho$ obtained using the proposed method.
The diagonal plots display the marginal posterior PDFs of $E$, $\nu$, and $\rho$. 
These PDFs are estimated using the Kernel Density Estimation method~\cite{wand1994kernel}.
Compared with the uniform prior, the posterior distribution of $\rho$ is the most concentrated, followed by that of $E$, suggesting that these parameters are more sensitive to the observations.
In contrast, the posterior of $\nu$ remains relatively flat, indicating limited influence from the observations.
The contour maps in the lower-triangular plots of the joint posterior PDF reveal the correlations among the parameters.

\subsection{Example 4: Thermal stress analysis of jet engine turbine blade}
\begin{figure}[htbp]
    \centering
    \subfigure[Finite element mesh]{
        \includegraphics[scale=0.5]{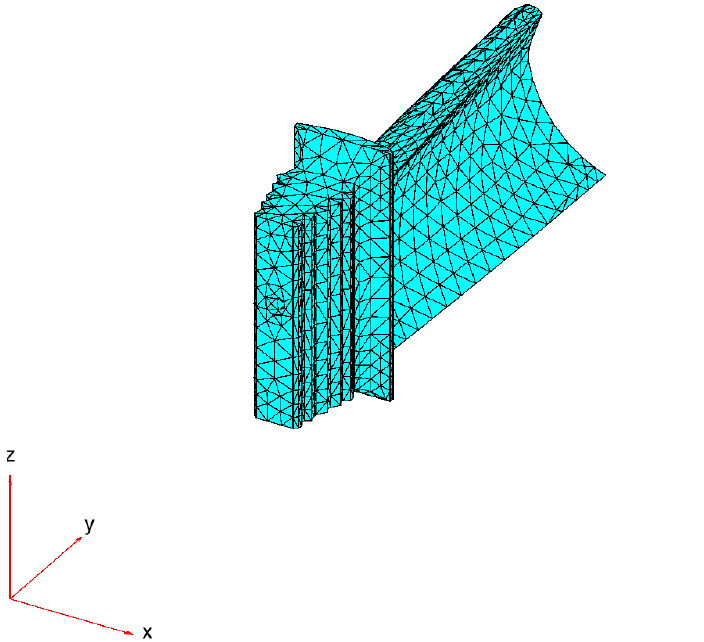} \label{fig:Ex4_mesh}
    }
    \subfigure[Von Mises stress]{
        \includegraphics[scale=0.35]{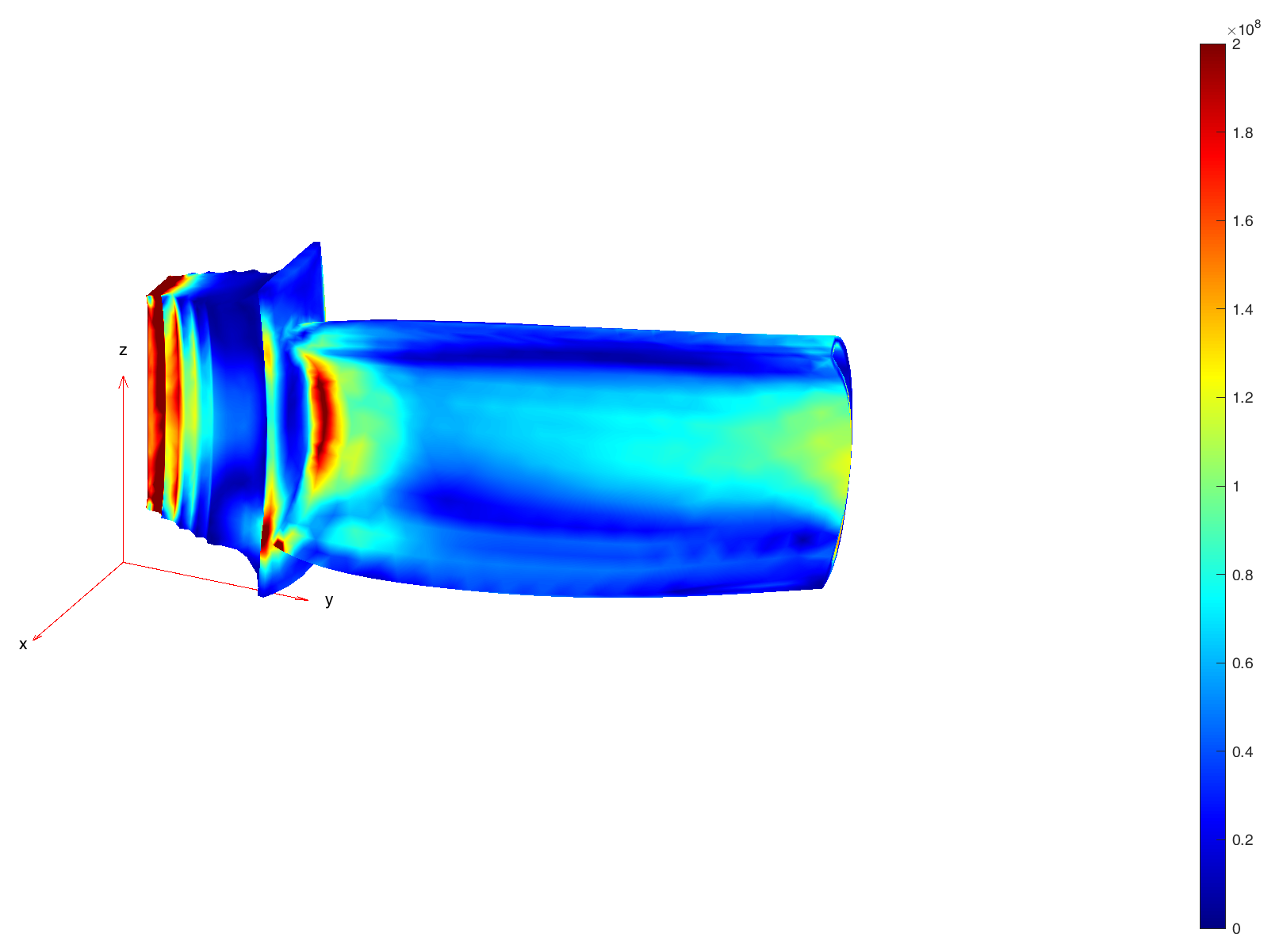} \label{fig:Ex4_heat}
    } 
    \caption{The finite element mesh and von Mises stress of jet engine turbine blade }
\label{fig:ex4_mesh_modes}
\end{figure}

The last example investigates a turbine blade structure, a critical component in jet engines.
This blade has internal cooling channels through which cool air flows to regulate the blade temperature within the range allowed by the material.
Turbine blades are typically made of nickel-based alloys for their excellent resistance to extreme thermal environments. 
However, prolonged exposure to high temperatures causes thermal expansion, which generates significant mechanical stresses at joints and leads to deformations on the order of millimeters.
To accurately capture these effects, thermal analysis of the blade is crucial.
The blades were modeled and analyzed using the Partial Differential Equation (PDE) Toolbox in MATLAB. 
The corresponding finite element mesh and von Mises stress distribution are shown in Figure ~\ref{fig:ex4_mesh_modes}.

The primary physical quantity of interest in this study is the maximum von Mises displacement of the blade, which is affected by several parameters, including Young's modulus ($E$), the coefficient of thermal expansion ($\gamma$), Poisson's ratio ($\nu$), pressure load on the pressure side ($P_1$), pressure load on the suction side ($P_2$), thermal conductivity ($K$) and the reference temperature ($T$).
The objective is to update the prior distributions of these parameters by considering the observed measurements of the blade's maximum displacement. 
The prior distributions for all parameters are summarized in Table~\ref{tab:ex_4_parameters}. 
Three displacement observations are used:  $u_{\mathrm{obs},1} = 0.0016\,\text{m}$, $u_{\mathrm{obs},2} = 0.0018\,\text{m}$, $u_{\mathrm{obs},3} = 0.0020\,\text{m}$.
Based on this setup, the likelihood function is defined as:
\begin{equation}
L(\boldsymbol{Y}_{\mathrm{obs}}|\boldsymbol{x}) = \exp\left(-\frac{1}{2\sigma_\epsilon^2} \sum_{i=1}^{3} \left(\frac{u^2(\boldsymbol{x})}{u_{\mathrm{obs},i}^2} - 1 \right)^2 \right),
\label{eq:likelihood_ex3}
\end{equation}
where $\sigma_\epsilon = 0.2$, and $u(\boldsymbol{x})$ denotes the model-predicted maximum von Mises displacement given the parameter vector $\boldsymbol{x}$.

\begin{table}[h]
\centering
\renewcommand{\arraystretch}{1.2} 
\caption{Prior distribution of uncertain parameters for Example 4}\label{tab:information_ex_4}
\begin{tabular}{cccccc}
\hline
Parameter & Physical meaning & Distribution type & Lower bound & Upper bound \\
\midrule
$E$ & Young's modulus (Pa) & Uniform &  $1.9295 \times 10^{11}$ &  $2.6105 \times 10^{11}$\\
$\gamma$ & Coefficient of thermal expansion (1/K) & Uniform & $ 1.0795 \times 10^{-5}$ & $1.4605 \times 10^{-5}$  \\
$\nu$ & Poisson's ratio & Uniform & 0.2295 & 0.3105 \\
$P_1$ & Pressure load on the pressure side (Pa) & Uniform & 9.7550 & 13.2250 \\
$P_2$ & Pressure load on the suction side (Pa) & Uniform &  $4.2500 \times 10^{5} $ &  $5.7500 \times 10^{5}$ \\
$K$ & Thermal conductivity (W/m/K) & Uniform &  $3.8250 \times 10^{5}$ &  $5.1750 \times 10^{5}$ \\
$T$ & Reference temperature (°C) & Uniform & 255 & 345 \\
\hline
\end{tabular}
\label{tab:ex_4_parameters}
\end{table}

Table~\ref{tab:ex_4_reults} presents a comparative analysis of the results obtained from BUS, TMCMC, and the proposed method. 
Due to the computational cost of FEA in this example, both BUS and TMCMC were executed only once, requiring 3.15 and 1.83 days, respectively. 
These results serve as benchmark references. 
In contrast, the proposed method ($n_0=10$) is significantly more efficient, taking only 0.18 hours per run. 
As shown in Table~\ref{tab:ex_4_reults}, the results produced by the proposed method are consistent with those from BUS and TMCMC and demonstrate low CoV.
Moreover, the proposed method requires significantly fewer model evaluations (only 88.25), demonstrating notable computational efficiency.

\begin{table}[h]
\centering
\renewcommand{\arraystretch}{1.2} 
\caption{Model updating results for Example 4}\label{tab:reults_ex_4}
\begin{tabular}{c@{\hspace{30pt}}cccccccccccc}
\hline
Method & BUS & TMCMC  & Proposed method  \\
\hline
{Time(s)} &   &     & 13297.08  \\
{$N_{call}$ (CoV)} & 10000 & 10000  & 88.25 (9.34\%)\\
{$\hat{c}$ (CoV)}  & 0.0763  &  0.0744   & 0.0737 (1.26\%) \\
{$\hat{\mu}_E$ (CoV)}  &$2.267 \times 10^{11}$  & $2.2625 \times 10^{11}$   & $2.2664 \times 10^{11}$ (0.15\%) \\
{$\hat{\mu}_{\gamma}$ (CoV)}  &$1.3657 \times 10^{-5}$   &  $1.3616 \times 10^{-5}$   &  $1.3641 \times 10^{-5}$ (0.06\%) \\
{$\hat{\mu}_{\nu}$ (CoV)}  &0.2703  & 0.2700  &  0.2699 (0.18\%)  \\
{$\hat{\mu}_{P_1}$ (CoV)}  &11.3610  & 11.3752 &  11.3666  (0.15\%) \\
{$\hat{\mu}_{P_2}$ (CoV)}  & $5.0702 \times 10^{5}$  & $5.0618 \times 10^{5}$ &  $5.0673\times 10^{5}$ (0.13\%) \\
{$\hat{\mu}_{K}$ (CoV)}  &$4.4474 \times 10^{5}$  &$4.4403 \times 10^{5}$ & $4.4477 \times 10^{5}$ (0.17\%) \\
{$\hat{\mu}_{T}$ (CoV)}  & 283.6655 & 285.5074 & 284.4551 (0.13\%) \\
{$\hat{\sigma}_E$ (CoV)} &$1.9631 \times 10^{10}$  & $1.8672 \times 10^{10}$ &  $1.9566 \times 10^{10}$ (0.75\%) \\
{$\hat{\sigma}_{\gamma}$ (CoV)} & $6.9606 \times 10^{-7}$ &  $6.7455 \times 10^{-7}$&  $7.0857 \times 10^{-7}$ (0.71\%) \\
{$\hat{\sigma}_{\nu}$ (CoV)} & 0.0234 & 0.0220 &  0.0234 (0.62\%)\\
{$\hat{\sigma}_{P_1}$ (CoV)} & 0.9868 & 0.9349  & 0.9930 (0.76\%) \\
{$\hat{\sigma}_{P_2}$ (CoV)} & $4.3234 \times 10^{4}$ &$4.0860 \times 10^{4}$  &  $4.3241 \times 10^{4}$ (0.55\%) \\
{$\hat{\sigma}_{K}$ (CoV)} & $3.9265 \times 10^{5}$ & $3.6431 \times 10^{4}$ &   $3.8866 \times 10^{5}$ (0.94\%)  \\
{$\hat{\sigma}_{T}$ (CoV)} & 20.9581 & 19.9603 &  21.3262  (1.25\%)\\
\hline
\end{tabular}
\label{tab:ex_4_reults}
\end{table}

\begin{figure}
    \centering
    \includegraphics[width=1\linewidth]{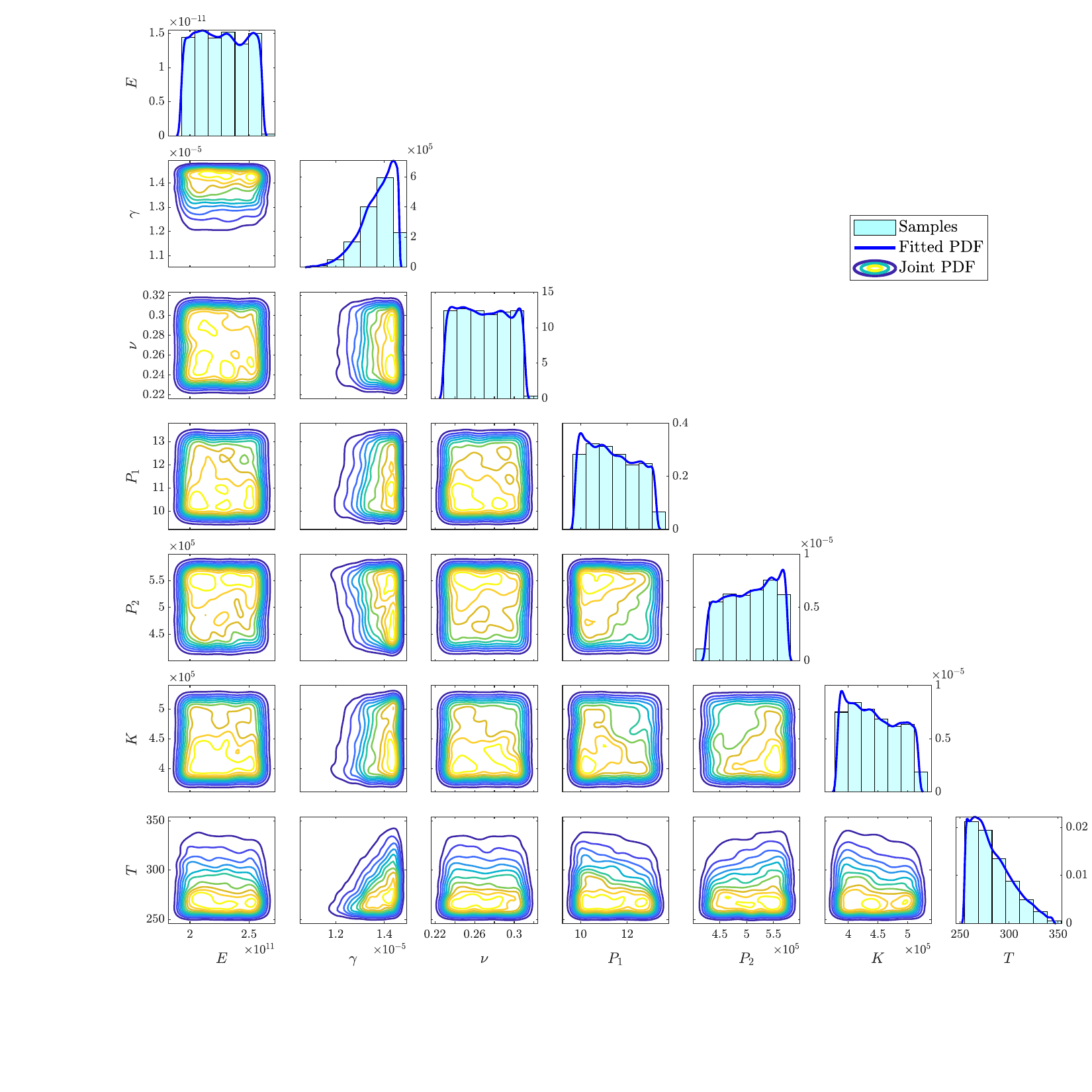}
    \caption{Posterior distributions of updated parameters in Example 4}
    \label{fig:Ex4_couter_plot}
\end{figure}

Fig.~\ref{fig:Ex4_couter_plot} displays the posterior distributions of the updated parameters obtained using the proposed method.
The diagonal plots show the marginal posterior PDFs, while the lower triangular subplots present contour plots of the joint posterior PDF.
Compared to their uniform priors, the posteriors of parameters such as $\gamma$ and $T$ are significantly more concentrated, indicating that these parameters are more sensitive to the observed data.
In contrast, the posterior of $E$, $\nu$ remains relatively flat, suggesting that they are less sensitive to the observations.
The joint distributions further reveal correlations between certain parameters, highlighting their statistical dependencies.

%
%
\section{Summary and conclusions}
\label{sec:Summary_conclusions}
In this study, a streamlined Bayesian active learning cubature (SBALC) is proposed for Bayesian model updating.
In this method, the log-likelihood function is approximated using Gaussian process (GP) regression in a streamlined Bayesian active learning way. 
Based on the posterior mean function of the log-likelihood function, a plug-in estimator for model evidence is introduced.
By deriving an upper bound on the absolute error between the posterior model evidence and the plug-in estimator, the epistemic uncertainty of the plug-in estimator can be quantified only through its upper and lower bounds, thereby eliminating the need for costly posterior GP sampling.
Building on this result, a novel stopping criterion is introduced to determine the termination of the iterative process, together with a new learning function for selecting the next most informative evaluation point.
As a result, SBALC requires only the posterior mean and variance functions of the GP to obtain a reliable estimate of model evidence, while simultaneously producing an approximate posterior distribution as a by-product.
The performance of the proposed method is demonstrated by four numerical examples, in comparison with several existing methods.
The results indicate that SBALC substantially reduces both the number of model evaluations and computational time, while maintaining high estimation accuracy and robustness.
This advantage is particularly evident in applications involving computationally intensive model analyses.

It is acknowledged that SBALC is currently limited to solving the likelihood function of moderate complexity. 
For highly non-linear or non-smooth likelihood function, particularly in high-dimensional settings, both the computational cost and the estimation error tend to increase significantly.
Future work will focus on enhancing SBALC by adopting more efficient stochastic simulation techniques to better address highly non-linear or non-smooth likelihood function, and by integrating advanced dimension-reduction strategies to cope with high-dimensional problems.
Moreover, efficiency can be further improved by evaluating multiple points per iteration, thereby enabling effective use of parallel computing resources.

\vspace{1em} %
\noindent\textbf{CRediT authorship statement} \\
\textbf{Pei-Pei Li}: Methodology, Software, Investigation, Writing-original draft, Writing-review \& editing, Funding acquisition. 
\textbf{Chao Dang}: Conceptualization, Methodology, Software, Investigation, Writing-review \& editing, Funding acquisition.
\textbf{Cristóbal H. Acevedo}: Methodology, Writing-review \& editing, Funding acquisition. 
\textbf{Marcos A. Valdebenito}: Conceptualization, Methodology, Investigation, Writing-review \& editing. 
\textbf{Matthias G.R. Faes}: Methodology, Investigation, Writing-review \& editing, Supervision, Funding acquisition.

\vspace{1em} 

\noindent\textbf{Acknowledgment} \\
The authors gratefully acknowledge the support of the Alexander von Humboldt Foundation for the postdoctoral grant of Pei-Pei Li, the Henriette Herz Scouting program (Matthias G.R. Faes), the State Key Laboratory of Disaster Reduction in Civil Engineering, Tongji University (Grant number SLDRCE24-02), the financial support of the German Research Foundation (DFG) for Chao Dang (Grant number 530326817), and the financial support of the dtec.bw-Digitalization and Technology Research Center of the Bundeswehr, which is funded by the European Union-NextGenerationEU.

%
%
\bibliographystyle{elsarticle-num}
\bibliography{Manuscript_SBALC.bib}
\end{document}